\documentclass[onecolumn]{IEEEtran}
\usepackage{cite}
\usepackage{amsmath,amssymb,amsfonts}
\usepackage{algorithmic}
\usepackage{graphicx}
\usepackage{epstopdf}
\usepackage{textcomp}
\usepackage{multirow}
\usepackage{booktabs}
\usepackage{amsmath}
\usepackage{float}
\usepackage{stfloats}
\usepackage{threeparttable}
\usepackage{booktabs}
\usepackage{tablefootnote}
\allowdisplaybreaks[4]
\def\BibTeX{{\rm B\kern-.05em{\sc i\kern-.025em b}\kern-.08em
    T\kern-.1667em\lower.7ex\hbox{E}\kern-.125emX}}
\begin{document}
\title{A Stable FDTD Subgridding Scheme with SBP-SAT for Transient Electromagnetic Analysis}
\author{Yu Cheng,~\IEEEmembership{Graduate Student Member,~IEEE,} Yuhui Wang, Hanhong Liu, Lilin Li, Xiang-Hua Wang, \\Shunchuan Yang, \IEEEmembership{Member,~IEEE}, and Zhizhang (David) Chen, \IEEEmembership{Fellow, IEEE}
\thanks{Manuscript received xxx; revised xxx.}
\thanks{This work was supported in part by the National Natural Science Foundation of China through Grant 61801010, 61671257, in part by Pre-Research Project through Grant J2019-VIII-0009-0170 and Fundamental Research Funds for the Central Universities \textit{(Corresponding author: Shunchuan Yang)}.}
\thanks{Y. Cheng, Y. Wang, and H. Liu are with the School of Electronic and Information Engineering, Beihang University, Beijing, 100083, China (e-mail: yucheng@buaa.edu.cn, yhwang\_0420@buaa.edu.cn, liu759753745@buaa.edu.cn).}
\thanks{L. Li and S. Yang are with Research Institute for Frontier Science and School of Electronic and Information Engineering, Beihang University, Beijing, China (e-mail: lll\_work@buaa.edu.cn, scyang@buaa.edu.cn).}
\thanks{X. H. Wang is with the School of Science, Tianjin University of Technology and Education, Tianjin, 300222, China (e-mail:  xhwang199@outlook.com.)}
\thanks{Z. Chen is currently with the College of Physics and Information Engineering, Fuzhou University, Fuzhou, Fujian. P. R. China, on leave from the Department of Electrical and Computer Engineering, Dalhousie University, Halifax, Nova Scotia, Canada B3H 4R2  (email: zz.chen@ieee.org). }
}

\maketitle

\begin{abstract}
We proposed a provably stable FDTD subgridding method for accurate and efficient transient electromagnetic analysis. In the proposed method, several field components are properly added to the boundaries of Yee's grid to make sure that the discrete operators meet the summation-by-parts (SBP) property. Then, by incorporating the simultaneous approximation terms (SATs) into the finite-difference time-domain (FDTD) method, the proposed FDTD subgridding method mimics the energy estimate of the continuous Maxwell’s equations at the semi-discrete level to guarantee its stability. Further, to couple multiple mesh blocks with different mesh sizes, the interpolation matrices are also derived. The proposed FDTD subgridding method is accurate, efficient, easy to implement and be integrated into the existing FDTD codes with only simple modifications. At last, three numerical examples with fine structures are carried out to validate the effectiveness of the proposed method.  
\end{abstract}


\section{Introduction}
\label{sec:introduction}
\IEEEPARstart{T}{he} finite-difference time-domain (FDTD) method is one of the most widely used numerical methods to solve Maxwell's equations in the time domain due to its powerful capability of dealing with complex media \cite{LARGE,MULTISCALE,FDTDCIRCUITS2,FDTDCIRCUITS3,FDTDCIRCUITS4}, easy implementation \cite{FDTDEMC,FDTDEMC2,FDTDEMC3,FDTDEMI,FDTDEMI2,FDTDEMI3}, and high parallel computational efficiency \cite{PARALLELFDTD1} \cite{PARALLELFDTD2}. However, it suffers from staircase error if complex structures are involved in the computational domain. 

The subgridding techniques can significantly improve the accuracy without sacrificing much of the FDTD efficiency since only locally fine meshes are used for geometrically delicate structures. Many subgridding techniques for the FDTD methods have been proposed in recent years, such as the wave equation based subgridding technique \cite{HOFDF}, the Huygens subgridding technique \cite{AFSBO}, the hybrid implicit-explicit (HIE)-FDTD based hybrid grid technique \cite{ANHIF}, the hybrid alternatively-direction-implicit (ADI)-FDTD method \cite{AIITI} and many others. In \cite{SFSWS} \cite{AS3FM}, a reduced order model was used in the subgridding FDTD method to extend the Courant-Friedrich-Levy (CFL) condition for efficiency improvement. A high-order smoothing technique with the complex nonstandard (CNS)-FDTD method was proposed to handle time interpolation between coarse and fine grids in \cite{IFCAF}. In \cite{SPSFS}, an unsymmetric FDTD subgridding algorithm with arbitrary refinement ratio was proposed to improve the accuracy. The adaptive mesh refinement (AMR)-FDTD method was proposed to model microwave integrated circuits \cite{FDTDCIRCUITS}. By incorporating the locally fine meshes, the accuracy can be significantly improved without severely increasing simulation time. 

However, the late-time stability of the subgridding techniques can not always be guaranteed and they may suffer from the so-called late-time instability \cite{STABILITYSUBGRIDDING}. Therefore, many efforts are made to develop theoretically stable subgridding methods to address this issue. In \cite{SPATIALFILTER}, a subgridding technique is theoretically made stable by decomposing solutions into stable and unstable modes and then filtering out unstable modes. However, the mode decomposition may be computationally intensive when electrically large or multiscale structures are involved. Recently, a decent dissipation theory is proposed to analyze the electromagnetic energy of the FDTD methods, and then was used to examine the stability of the subgridding techniques in \cite{AS3FM} \cite{ADSTF}. 

Local grid refinement techniques are also apllied in other research fields.  One such method is the summation-by-parts simultaneous approximation term (SBP-SAT) technique \cite{SPBEARLY}, which was originally proposed to develop the provably long-time stable finite-difference methods for the Euler and Navier-Stokes equations \cite{SPB} \cite{SPBNS}, wave propagation \cite{SFSWS} \cite{SPBEARLYWAVEMCOLLOCATED} \cite{SBPWAVE} and other applications. Interested readers are referred to two comprehensive review papers \cite{ROSSF} \cite{ROSOW} for more technical details. The SBP-SAT techniques are reported to solve the Maxwell’s equations in  the two-dimensional space\cite{SPBEARLYWAVEMCOLLOCATED} \cite{ESSMF}. However, those previous works are only applicable when electric and magnetic field components are collocated on each grid point. With the FDTD method of Yee's grid, electric-field and magnetic-field nodes are collocated but staggered and interlaced by one- half -cell, apart for better numerical dispersion and accuracy \cite{ESAHF}. Therefore, many efforts are also made to develop the SBP-SAT methods with staggered grids in acoustic applications \cite{ESAHF,CDSAU, PROJECTSBP}.

Inspired by the work \cite{ESAHF}, we proposed a theoretically stable FDTD subgridding method in conjugation with the SBP-SAT technique in this paper. With carefully adding field components on the boundaries of the Yee's grid, the discrete operators automatically satisfy the SBP property, and the electromagnetic energy in the computational domain is fully determined by field components on the boundaries. By incorporating the SAT to weakly enforce the boundary conditions with multiple mesh blocks with different mesh sizes, the stability of the proposed SBP-SAT FDTD method is rigorously proved and theoretically guaranteed. Therefore, it is numerically stable in the long-time simulations.

This paper proposes a stable method with the SBP-SAT methods for locally refined grids.  In comparison with the work presented in \cite{SPBEARLYWAVEMCOLLOCATED} \cite{ESSMF}, several advances have been made in this paper. First, similar to the traditional FDTD method, staggered grids are used in the proposed FDTD method. Secondly, only field nodes are added to the boundaries of the original Yee's grid. Thirdly, an analysis of the stability condition of the proposed SBP-SAT FDTD method is presented. The interpolation matrices to couple multiple mesh blocks are analytically derived. Therefore, the late-time stability of the proposed FDTD subgridding method is theoretically guaranteed.

The contributions of this paper are of three folds.
\begin{enumerate}
\item The SBP-SAT FDTD method with staggered grids is proposed to solve Maxwell's equation in the two-dimensional transverse magnetic (TM) mode. The grids are constructed by adding electric field nodes at the four corners and at the edge centers of all the boundary cells. Compared with the original FDTD method, the time-marching formulations in the inner regions are kept unchanged and are only modified near the boundaries. Therefore, only a few modifications are needed to be incorporated into the existing FDTD codes. Since the explicit leapfrog scheme is used to update fields in the time domain, the time steps of the proposed SBP-FDTD method are also constrained by cell sizes and boundary conditions in the same way as the traditional FDTD method. The CFL condition of the SPB-SAT FDTD method is analytically derived.  

\item Late-time stability can be guaranteed.

\item The interpolation matrices between two mesh blocks with different mesh sizes are derived. The SBP-SAT technique is incorporated into the proposed FDTD subgridding method to handle the boundary conditions at the interfaces of different mesh blocks, and the interpolation rules between two mesh blocks are developed to satisfy the norm compatible conditions.
\end{enumerate}

This paper is organized as follows. In Section II, detailed formulations of the proposed SBP-SAT FDTD method and the stability condition proof in a single mesh block are presented. In Section III, the formulations for connecting multiple mesh blocks are shown. In Section IV, the CFL condition of the proposed SBP-SAT FDTD method is verified. In Section V, three numerical examples are carried out to validate the accuracy and efficiency of the proposed SBP-SAT FDTD method and the FDTD subgridding method with the SBP-SAT technique. At last, we draw some conclusions in Section VI.

\section{THE FDTD METHOD WITH THE SBP-SAT IN A SINGLE MESH BLOCK}
\subsection{Problem Configurations and Notations}
 As shown in Fig. 1, a local fine mesh surrounded by a coarse mesh is considered in this paper. In this configuration, coarse orthogonal structural meshes are used in the exterior domain, and the local fine mesh is embedded into coarse meshes to model geometrically fine structures. 

The following notations are used in our derivation. A hollow character denotes a matrix, and its subscript denotes the corresponding grid. A bold character, e.g., ${\bf{A}}$, is a column vector, and its subscript, is the dreiction; e.g., ${\bf{E}}_z$, denotes the electric field $E$ in the z-direction.  $\otimes$ is the Kronecker product of two matrices of arbitrary sizes.

\begin{figure} \label{first}	
	\centerline{\includegraphics[scale=0.5]{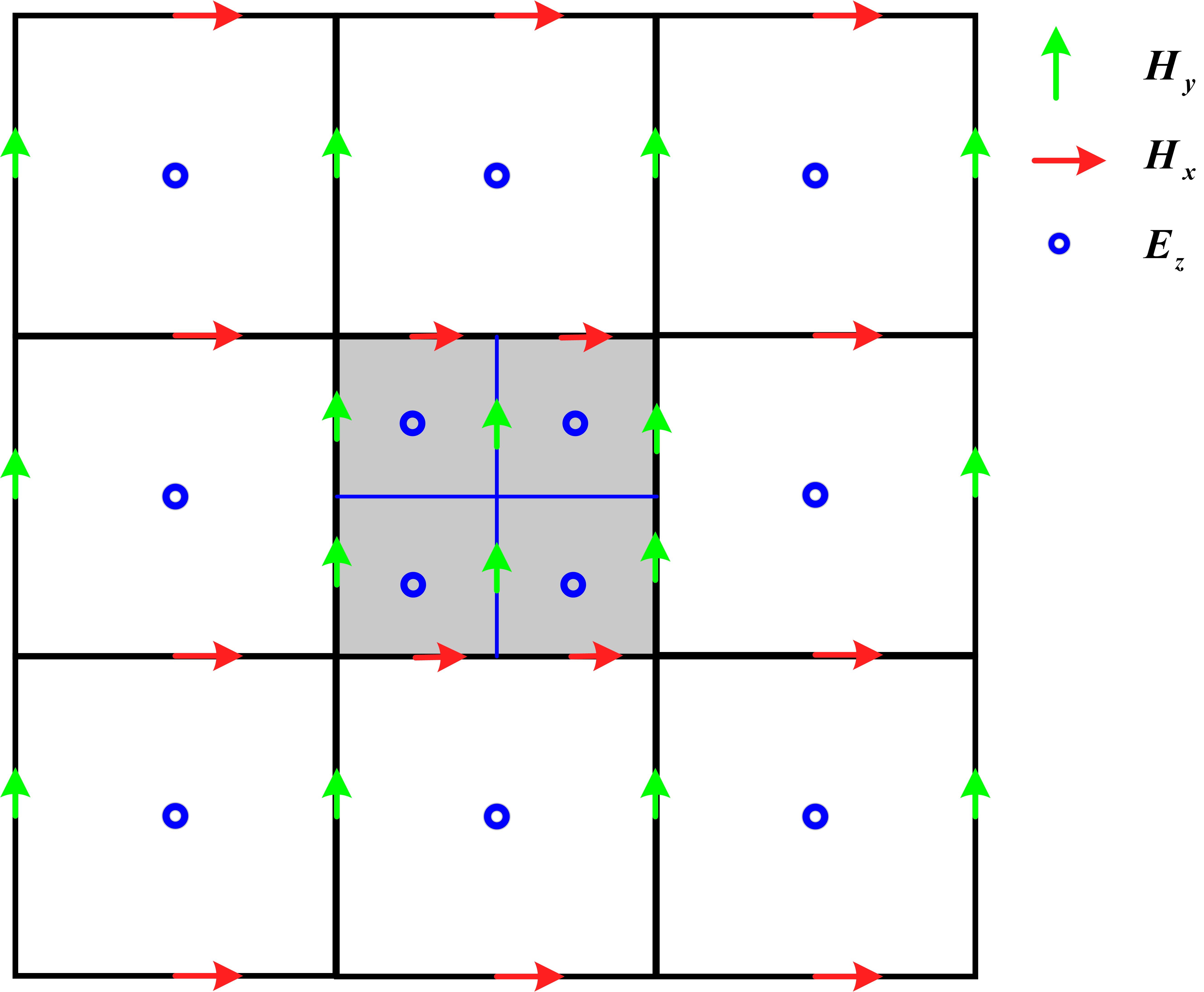}}	
	\caption{A local fine mesh used to model the geometrically fine structures in the inner region is embedded into coarse meshes in the two-dimensional TM mode.
	}
\end{figure}

The following two-dimensional eqautions are condiered: 
\begin{subequations} \label{maxwell} 
\begin{align}
&\frac{{\partial {H_x}}}{{\partial t}} = - \frac{1}{\mu} \frac{{\partial {E_z}}}{{\partial y}},\\
&\frac{{\partial {H_y}}}{{\partial t}} = \frac{1}{\mu} \frac{{\partial {E_z}}}{{\partial x}},\\
&\frac{{\partial {E_z}}}{{\partial t}} = \frac{1}{\varepsilon} \left(\frac{{\partial {H_y}}}{{\partial x}} - \frac{{\partial {H_x}}}{{\partial y}} \right),
\end{align}
\end{subequations}
where ${{H_x}}$ and ${{H_y}}$ are the magnetic fields in the $x$ and $y$ direction, respectively, and ${{E_z}}$ denotes the electric field in the $z$ direction. $\varepsilon$ and $\mu$  are the permittivity and permeability of the medium, respectively.

\subsection{Field Component Locations in A Single Mesh Block}
\begin{figure}  
		{\includegraphics[scale=0.6]{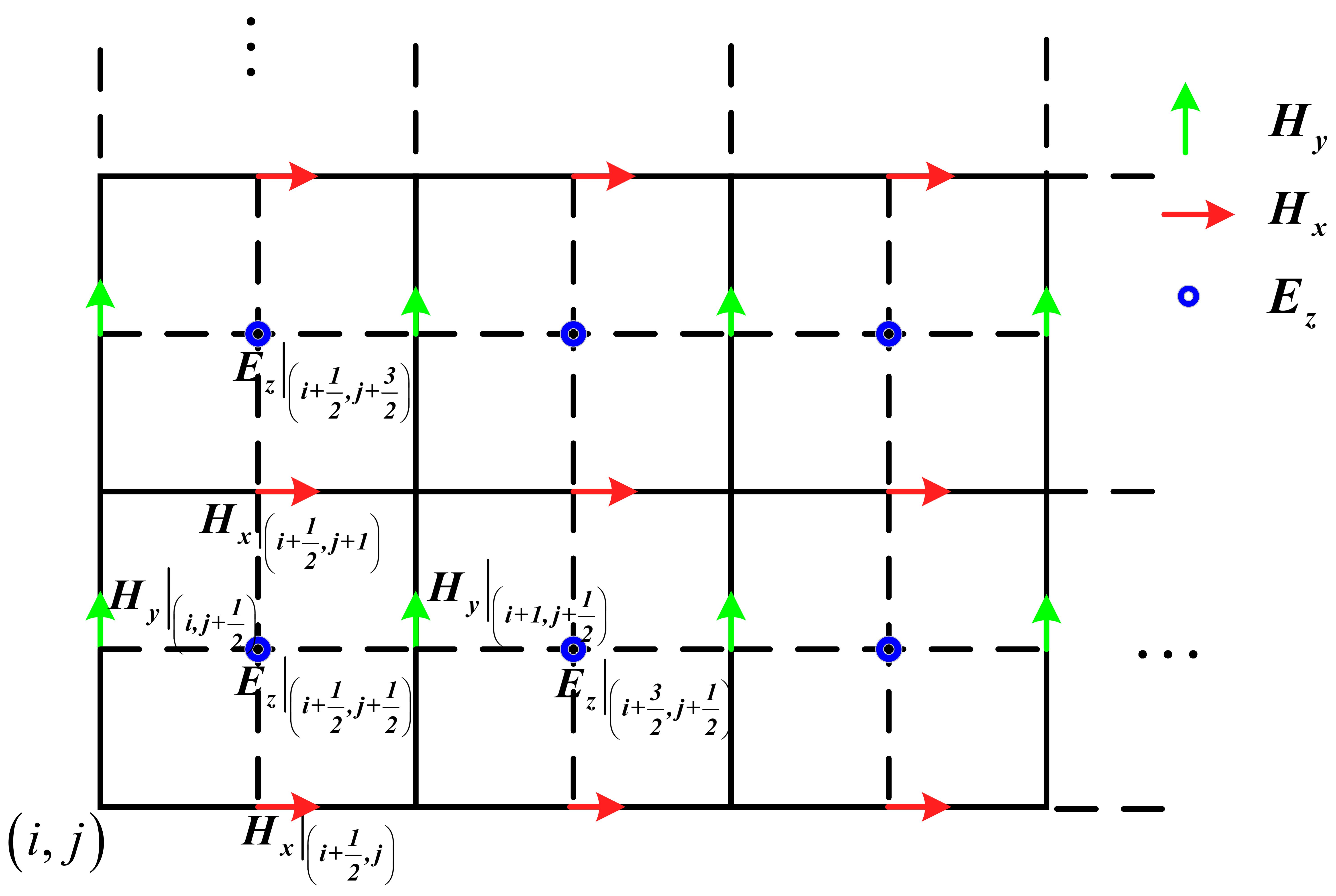}}
	 	\centerline{(a)}
 	 {\includegraphics[scale=0.6]{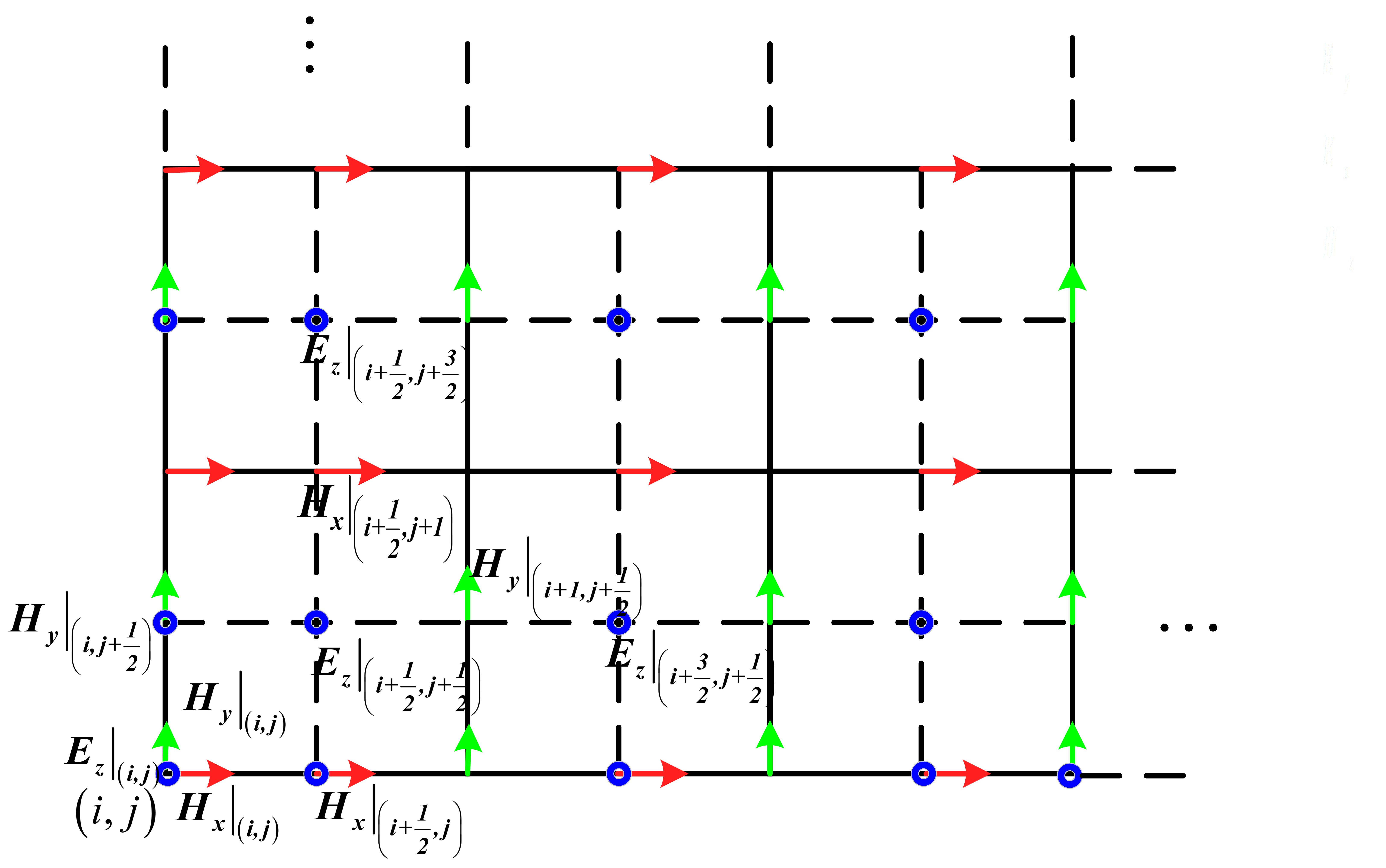}}
		\centerline{(b)}
	\centering
	\caption{(a) The TM-mode node distribution in the two-dimensional Yee's grid, (b) the staggered grids used in the proposed SBP-SAT FDTD method, where four electric nodes are added at the four corners and at the middle of boundary edges of the computational domain.}
	\label{FieldLocations}
\end{figure}

The electric and magnetic field nodes in Yee's grid are interlaced with each other. Each electric field node is located at the cell center, and surrounded by four magnetic field nodes on the boundaries of each cell, and vice versa. As shown in Fig. \ref{FieldLocations}(a), ${{E_z}}$ nodes are located at the centers of grid cells, and ${{H_x}}$ and ${{H_y}}$ nodes are located in the middle of each cell edge in the $x$ and $y$ direction, respectively. In contrast to the Yee's grid in Fig. \ref{FieldLocations}(a), we keep field nodes unchanged strictly inside the computational domain, and add additional ${{E_z}}$ nodes at the four corners, and at the middle of all the boundary edges of cells in the computational domain, as shown in Fig. \ref{FieldLocations}(b). In the following derivation, uniform meshes with the cell size $h$ and ${N_x}$, ${N_y}$ cells in the $x$ and $y$ directions are used. Therefore, the overall count of cells is ${N_x} \times {N_y}$ in the computational domain.  

\subsection{The Staggered SBP Operators}
We first briefly review the one-dimensional SBP operator introduced in \cite{ESSMF}. The SBP operators in the discrete grids are used to mimic the continuous operators and make the energy of the discrete system is fully determined by the field on the boundaries. Two staggered grids, ${\bf{x}}_- = {\left[ {{x_0},{x_{1/2}}, \ldots, {x_{n-1/2}}, {x_n}}\right]^T}$, ${\bf{x}}_{\rm{ + }} = {\left[ {{x_0},{x_1}, \ldots, {x_n}}\right]^T}$, are used to define two one-dimensional SBP operators. Note that the coordinate of grid point is $x_i = ih$.

To well define the SBP operators, they should satisfy the following accuracy relationship on ${\bf{x}}_-$ and ${\bf{x}}_{\rm{ + }}$, which can be expressed as
\begin{align} \label{Dop} 
{{\mathbb{D_ +}}{\bf{x}}_ - ^k = k{\bf{x}}_ + ^{k - 1},~{\mathbb{D_ -}{\bf{x}}_ + ^k = k{\bf{x}}_ - ^{k - 1}}},  \qquad  k=0,1
\end{align}
where $\mathbb{D}_+$ and $\mathbb{D}_-$ with dimension of ${N_{+ }} \times {N_{- }}$ and ${N_{- }} \times {N_{+ }}$ denote the discrete partial differential operators to approximate their continuous counterparts in the SBP-FDTD method. It should be noted that when $k = 0$, ${\bf{x}}_ - ^{- 1} = \bf{0}$ and ${\bf{x}}_ + ^{- 1} = \bf{0}$. $\mathbb{D}_+$ and $\mathbb{D}_-$ are further defined as 
\begin{align} \label{Dzf} 
{{\mathbb{D}}_+ = {{\mathbb{P}}_+^{{\bf{ - 1}}}}{\mathbb{Q}}}_+,~{{\mathbb{D}}_- = {\mathbb{P}_-^{{\bf{ - 1}}}}{\mathbb{Q}}_-},
\end{align} 
where $\mathbb{P}_-$ and $\mathbb{P}_+$ with dimension of ${N_{- }} \times {N_{- }}$ and ${N_{+ }} \times {N_{+ }}$ are the positive diagonal matrices defined on ${\bf{x}}_-$ and ${\bf{x}}_+$. Their entities denote the Gaussian weights associated with the corresponding field nodes. Therefore, we have  
\begin{align}\label{Energy}
{\left\| {\bf{x }} \right\|^2 = {{\bf{x }}^T}\mathbb{P}{\bf{x }} \approx \int {{x^2}dl} } ,
\end{align}
where $\left\| {\bf{. }} \right\|^2$  denote the two-norm of a column vector $\bf{x}$. The energy estimate of $x$ is approximated by $\left\| {\bf{x}} \right\|^2$.

$\mathbb{Q}_-$ and $\mathbb{Q}_+$ satisfy the SBP property if it meets the following relationship
\begin{align} \label{Q} 
{{{\mathbb{Q}}_ + } + {\mathbb{Q}}_ - ^T = {\mathbb{B}}  } ,
\end{align}
where ${\mathbb{B}} = diag\left[ { - 1,0 \ldots 0,1} \right]$. Obviously, we can obtain the following identity
\begin{align} \label{B} 
{{\bf {x}} ^T}{\mathbb{B}} {\bf {x}} = {x_n}{x_n} - {x_0}{x_0} .
\end{align}
Through (\ref{maxwell}), (\ref{Dzf})-(\ref{B}), the energy in the computational domain is fully determined by the boundary values. Therefore, it can be used to derive a stable FDTD method in the long-time simulations.

\subsection{The Proposed SBP-SAT FDTD Method for A Single-Block Mesh }   
In the SBP-SAT FDTD method, the partial differential operators, $\frac{\partial}{\partial x}$ and $\frac{\partial}{\partial y}$, in (\ref{maxwell}) are approximated by the corresponding discrete counterparts. In our implementation, the second-order central-difference scheme is used in the spatial domain. The semi-discrete formulations of (\ref{maxwell}) can be expressed as
\begin{subequations} \label{sbpequ} 
\begin{align}
	&{\frac{{d{{\bf{H}}_x}}}{{dt}}  + \frac{1}{\mu }\left( {{{\mathbb{I}}_{x - }} \otimes {{\mathbb{D}}_{y + }}} \right){{\bf{E}}_z} = {\bf{0}}}, \label{SEMI-DISCRETEEQU} \\ 
	&{\frac{{d{{\bf{H}}_y}}}{{dt}}   - \frac{1}{\mu } \left( {{{\mathbb{D}}_{x + }} \otimes {{\mathbb{I}}_{y - }}} \right){{\bf{E}}_z} = {\bf{0}}}, \label{SEMI-DISCRETEEQU2} \\
	&{\frac{{d{{\bf{E}}_z}}}{{dt}}  - \frac{1}{\varepsilon } \left( {{{\mathbb{D}}_{x - }} \otimes {{\mathbb{I}}_{y - }}} \right){{\bf{H}}_y} + \frac{1}{\varepsilon }\left( {{{\mathbb{I}}_{x - }} \otimes {{\mathbb{D}}_{y - }}} \right){{\bf{H}}_x} = {\bf{0}}},\label{SEMI-DISCRETEEQU3} 
\end{align}
\end{subequations}
where ${{\bf{H}}_x}$ is the column vector collecting all the $H_x$ nodes in a row-by-row manner in the computational domain, and similarly for ${{\bf{H}}_y}$ and ${{\bf{E}}_z}$. ${\mathbb{D}}_{x-}$, ${\mathbb{D}}_{x+}$, ${\mathbb{D}}_{y-}$ and ${\mathbb{D}}_{y+}$ are discrete partial differential matrices with dimensions of ${N_{x - }} \times {N_{x + }}$, ${N_{x + }} \times {N_{x - }}$, ${N_{y - }} \times {N_{y + }}$ and ${N_{y + }} \times {N_{y - }}$, respectively, where ${N_{x_-}}$, ${N_{x_+}}$, ${N_{y_-}}$, and ${N_{y_+}}$ denote the numbers of nodes on ${{\bf x}_-}$, ${{\bf x}_+}$, ${{\bf y}_-}$, and ${{\bf y}_+}$, respectively. It should be noted that ${{\bf y}_-}$, and ${{\bf y}_+}$ are the one-dimensional staggered grids in the $y$ direction.   ${\mathbb{I}}_{x-}$ and ${\mathbb{I}}_{y-}$ are identity matrices with dimensions of ${N_{x - }} \times {N_{x - }}$ and ${N_{y - }} \times {N_{y - }}$, respectively.

Assume that the perfect electric conductor (PEC) boundary conditions are used in the following derivation. In the proposed SBP-SAT FDTD method, the boundary conditions are weakly enforced through the SAT technique. In essence, the SAT technique adds penalty terms to (\ref{SEMI-DISCRETEEQU})-(\ref{SEMI-DISCRETEEQU3}), which are similar to the numerical flux in the discontinuous Galerkin finite element method (DG-FEM) \cite{ROSSF} \cite{ROSOW}. Therefore, after adding the corresponding SAT terms to (\ref{SEMI-DISCRETEEQU})-(\ref{SEMI-DISCRETEEQU3}), they can be rewritten as 
\begin{subequations} \label{sbpsatequ} 
\begin{align}
\frac{{d{{\bf{H}}_x}}}{{dt}} &+ \frac{1}{\mu }\left( {{{\mathbb{I}}_{x - }} \otimes {{\mathbb{D}}_{y + }}} \right){{\bf{E}}_z} \notag\\
&={\sigma _{s_0}}{\left( {{{\mathbb{P}}_{x - }} \otimes {{\mathbb{P}}_{y + }}} \right)^{ - 1}}{\mathbb{R}}_{{H_x}_S}^T{{\mathbb{P}}_{x - }}{{\bf{E}}_{z_S}}   \label{ZZb}\\
&+ {\sigma _{n_0}}{\left( {{{\mathbb{P}}_{x - }} \otimes {{\mathbb{P}}_{y + }}} \right)^{ - 1}}{\mathbb{R}}_{{H_x}_N}^T{{\mathbb{P}}_{x - }}{{\bf{E}}_{z_N}} , \notag\\
\frac{{d{{\bf{H}}_y}}}{{dt}} &- \frac{1}{\mu } \left( {{{\mathbb{D}}_{x + }} \otimes {{\mathbb{I}}_{y - }}} \right){{\bf{E}}_z}  \notag\\
&= {\sigma _{e_0}}{\left( {{{\mathbb{P}}_{x + }} \otimes {{\mathbb{P}}_{y - }}} \right)^{ - 1}}{\mathbb{R}}_{{H_y}_E}^T{{\mathbb{P}}_{y - }}{{\bf{E}}_{z_E}}  \label{ZZc}\\
& + {\sigma _{w_0}}{\left( {{{\mathbb{P}}_{x + }} \otimes {{\mathbb{P}}_{y - }}} \right)^{ - 1}}{\mathbb{R}}_{{H_y}_W}^T{{\mathbb{P}}_{y - }}{{\bf{E}}_{z_W}} , \notag\\
\frac{{d{{\bf{E}}_z}}}{{dt}} & - \frac{1}{\varepsilon }  \left( {{{\mathbb{D}}_{x - }} \otimes {{\mathbb{I}}_{y - }}} \right){{\bf{H}}_y} +  \frac{1}{\varepsilon }  \left( {{{\mathbb{I}}_{x - }} \otimes {{\mathbb{D}}_{y - }}} \right){{\bf{H}}_x}  \notag\\
&={\sigma _{s_1}}{\left( {{{\mathbb{P}}_{x - }} \otimes {{\mathbb{P}}_{y - }}} \right)^{ - 1}}{\mathbb{R}}_{{E_z}_S}^T{{\mathbb{P}}_{x - }}{{\bf{E}}_{z_S}}  \notag\\
&+{\sigma _{n_1}}{\left( {{{\mathbb{P}}_{x - }} \otimes {{\mathbb{P}}_{y - }}} \right)^{ - 1}}{\mathbb{R}}_{{E_z}_N}^T{{\mathbb{P}}_{x - }}{{\bf{E}}_{z_N}}\label{ZZa}\\
&+ {\sigma _{e_1}}{\left( {{{\mathbb{P}}_{x - }} \otimes {{\mathbb{P}}_{y - }}} \right)^{ - 1}}{\mathbb{R}}_{{E_z}_E}^T{{\mathbb{P}}_{y - }}{{\bf{E}}_{z_E}}  \notag\\
&+{\sigma _{w_1}}{\left( {{{\mathbb{P}}_{x - }} \otimes {{\mathbb{P}}_{y - }}} \right)^{ - 1}}{\mathbb{R}}_{{E_z}_W}^T{{\mathbb{P}}_{y - }}{{\bf{E}}_{z_W}} . \notag
\end{align}
\end{subequations}
Here the terms on the right hand of (\ref{ZZb})-(\ref{ZZa}) are the SATs to weakly enforce the PEC boundary conditions, and ${\bf{E}}_{z_S}^{}$, ${\bf{E}}_{z_N}^{}$, ${\bf{E}}_{z_E}^{}$, ${\bf{E}}_{z_W}^{}$ are four column vectors collecting all the ${{E}}_{z}^{}$ nodes on the south, north, east and west boundaries, respectively. They can be calculated by ${{\bf{E}}_{{z_S}}} = {{\mathbb{R}}_{{E_z}_S}}{{\bf{E}}_z}$, ${{\bf{E}}_{{z_N}}} = {{\mathbb{R}}_{{E_z}_N}}{{\bf{E}}_z}$, ${{\bf{E}}_{{z_E}}} = {{\mathbb{R}}_{{E_z}_E}}{{\bf{E}}_z}$, and ${{\bf{E}}_{{z_W}}} = {{\mathbb{R}}_{{E_z}_W}}{{\bf{E}}_z}$, respectively. In addition, we have ${{\mathbb{R}}_{{E_z}_S}} = {{\mathbb{I}}_{x-}} \otimes {\bf{e}}_{{y_0} - }^T$, ${{\mathbb{R}}_{{E_z}_N}} = {{\mathbb{I}}_{x-}} \otimes {\bf{e}}_{{y_N}-}^T$, ${{\mathbb{R}}_{{E_z}_W}} = {{\mathbb{I}}_{y -}} \otimes {\bf{e}}_{{y_0} - }^T$, ${{\mathbb{R}}_{{E_z}_E}} = {{\mathbb{I}}_{y -}} \otimes {\bf{e}}_{{y_N}-}^T$, where ${{\bf{e}}_{{y_0} - }} = {\left[ {\begin{array}{*{20}{c}}
		1&0&0&{...}&0
		\end{array}} \right]^T}$ with dimension of ${N_{y - }} \times 1$, and ${{\bf{e}}_{{y_N} - }} = {\left[ {\begin{array}{*{20}{c}}
		0&0&0&{...}&1
		\end{array}} \right]^T}$ with dimension of ${N_{y - }} \times 1$.  ${\mathbb{P}}_{y+}$, ${\mathbb{P}}_{x+}$, ${\mathbb{P}}_{x-}$ and ${\mathbb{P}}_{y-}$ are the diagonal norm matrices with dimensions of ${N_{y + }} \times {N_{y + }}$, ${N_{x + }} \times {N_{x + }}$, ${N_{x - }} \times {N_{y - }}$ and ${N_{y - }} \times {N_{y - }}$, respectively.  ${\sigma _{s_0}}$, ${\sigma _{w_0}}$, ${\sigma _{n_0}}$, ${\sigma _{e_0}}$, ${\sigma _{s_1}}$, ${\sigma _{w_1}}$, ${\sigma _{n_1}}$, and ${\sigma _{e_1}}$ are free parameters to guarantee the stability of the proposed SBP-SAT FDTD method, which can be determined in the following.

The total electromagnetic energy equals the sum of the energy of ${{\bf{H}}_x}$, ${{\bf{H}}_y}$ and ${{\bf{E}}_z}$ in the computational domain. According to (\ref{Energy}), the total electromagnetic energy ${\cal E}$ is expressed as
  \begin{align} \label{energysbp} 
 &{\cal E}  = \frac{1}{2}{{\bf{E}}_z}^T\left( {{{\mathbb{P}}_{x - }} \otimes {{\mathbb{P}}_{y - }}} \right){{\bf{E}}_z} \\
	&+ \frac{1}{2}{{\bf{H}}_y}^T\left( {{{\mathbb{P}}_{x + }} \otimes {{\mathbb{P}}_{y - }}} \right){{\bf{H}}_y} + \frac{1}{2}{{\bf{H}}_x}^T\left( {{{\mathbb{P}}_{x - }} \otimes {{\mathbb{P}}_{y + }}} \right){{\bf{H}}_x}   .\notag
   \end{align} 
By taking its partial derivative with respect to time, we obtain
\begin{align} \label{energyd} 
\frac{{d{\cal E}}}{{dt}} &= {\sigma _{s_1}}{{\bf{E}}_{z_S}}^T{{\mathbb{P}}_{x - }}{{\bf{E}}_{z_S}} + \left( { {\sigma _{s_0}}+1} \right){{\bf{E}}_{z_S}}^T{{\mathbb{P}}_{x - }}{{\bf{H}}_{x_S}}\notag\\
&+ {\sigma _{n_1}}{{\bf{E}}_{z_N}}^T{{\mathbb{P}}_{x - }}{{\bf{E}}_{z_N}} + \left( {{\sigma _{n_0}}-1} \right){{\bf{E}}_{z_N}}^T{{\mathbb{P}}_{x - }}{{\bf{H}}_{x_N}}\\
&+ {\sigma _{w_1}}{{\bf{E}}_{z_W}}^T{{\mathbb{P}}_{y - }}{{\bf{E}}_{z_W}} + \left( { {\sigma _{w_0}}+1} \right){{\bf{E}}_{z_W}}^T{{\mathbb{P}}_{y - }}{{\bf{H}}_{y_W}}\notag\\
&+ {\sigma _{e_1}}{{\bf{E}}_{z_E}}^T{{\mathbb{P}}_{x - }}{{\bf{E}}_{z_E}} + \left( { {\sigma _{e_0}}-1} \right){{\bf{E}}_{z_E}}^T{{\mathbb{P}}_{y - }}{{\bf{H}}_{y_E}}.\notag
\end{align}
When there is no dissipation of energy in the computational domain, ${{d{\cal E}}}/{{dt}} = 0$ should be satisfied. Therefore, to ensure that the proposed SBP-SAT FDTD method is stable in the long-time simulations, one intrinsic option is that ${\sigma _{s_0}} = {\sigma _{w_0}} =  - 1$, ${\sigma _{n_0}} = {\sigma _{e_0}} = 1$, and ${\sigma _{s_1}} = {\sigma _{n_1}} = {\sigma _{e_1}} = {\sigma _{w_1}} = 0$, respectively.
 
\section{STABLE COUPLING INTERACTIONS FOR MULTIPLE MESH BLOCKS}
\subsection{Stable Coupling Interactions between Multiple Mesh Blocks} 
\begin{figure}  
	\centerline{\includegraphics[scale=0.5]{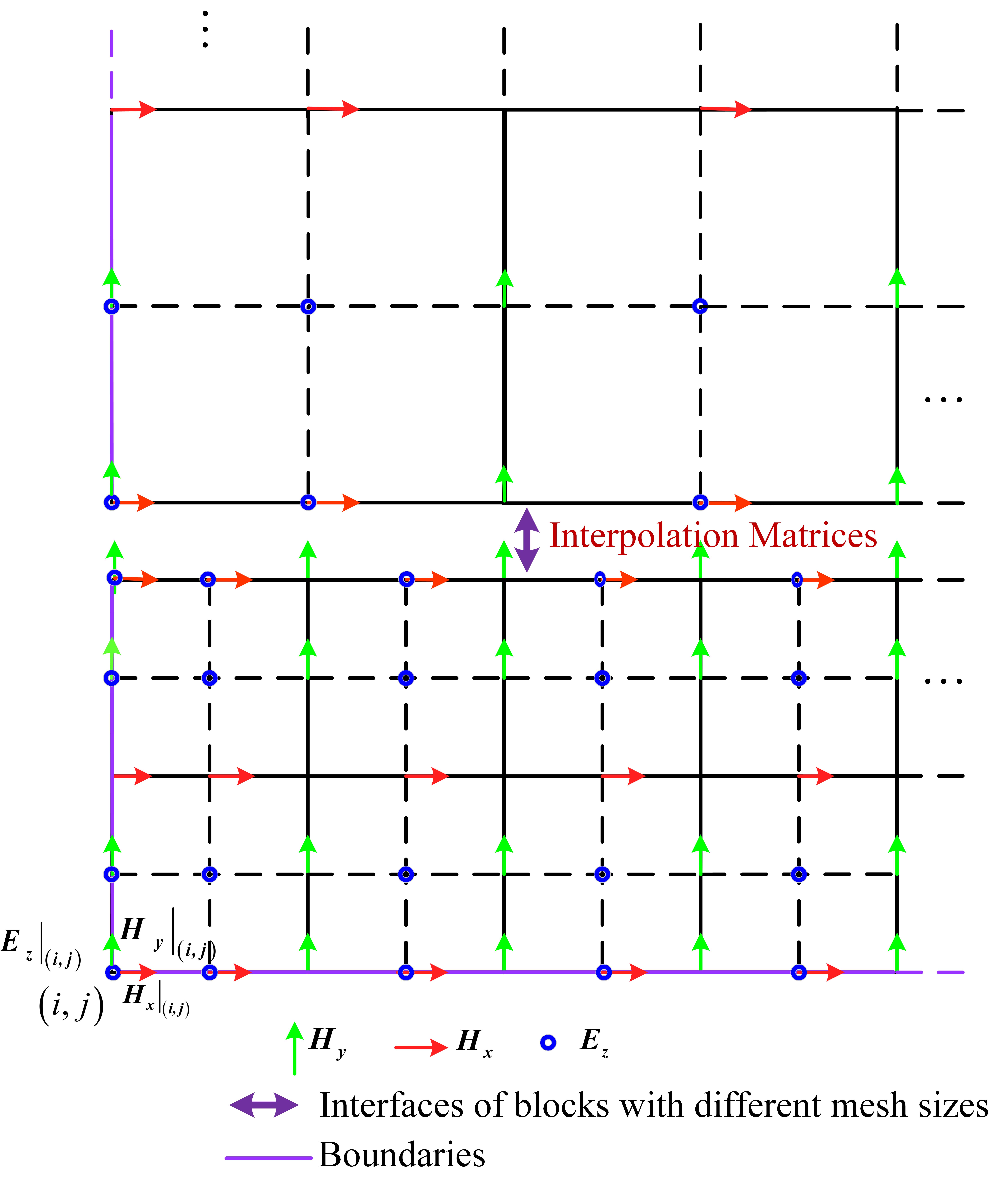}}	
	\caption{Two mesh blocks with different mesh sizes are connected with each other in the $y$ direction, and the cell size ratio is 2:1.}
	\label{SUBGRIDDING}	
\end{figure}

In this subsection, the SBP-SAT FDTD method for stable coupling interactions between two mesh blocks with different mesh sizes is derived. As shown in Fig. \ref{SUBGRIDDING}, two mesh blocks are connected with each other in the $y$ direction, and the cell size ratio is fixed as 2:1 for better visualization. However, any other cell size ratios are also possible if the appropriate interpolation matrices are defined. The southern boundary of the top domain is connected with the northern boundary of the bottom domain, and other boundaries keep the PEC boundary conditions unchanged. The node distribution in the two domains are the same as those in Fig. \ref{FieldLocations}(b). To distinguish the quantities in different domains, characters with a $\ \widehat{}\ $, e.g., ${{{\bf{\widehat E}}}_z}$, denotes that it is defined in the bottom domain and characters without the $\ \widehat{} \ $ for the top domain.
 
To accurately couple the two domains in Fig. \ref{SUBGRIDDING}, the boundary conditions at the interfaces are expressed as ${{\bf{E}}_z}_{_s} = {{\bf{\widehat E}}_{{z_n}}}$ and ${{\bf{H}}_x}_{_s} = {{\bf{\widehat H}}_{{x_n}}}$. In the following derivation, we assume that the PEC boundary conditions in the top and bottom domains are properly dealt with the SATs as mentioned in the previous section. Therefore, we need to consider the boundary conditions at the interfaces of the two mesh blocks through the SAT technique. The SBP-SAT FDTD formulations in the top domain are expressed as  
\begin{subequations}\label{upper} 
\begin{align}
& \,\frac{{d{{\bf{H}}_x}}}{{dt}} + \frac{1}{\mu } \left( {{{\mathbb{I}}_{x - }} \otimes {{\mathbb{D}}_{y + }}} \right){{\bf{E}}_z} =\notag\\
& \,\,\,\,\,\,\,\,\,\,\,\,\,\,\,\, {\sigma _{{n_1}}}{\left( {{{\mathbb{P}}_{x - }} \otimes {{\mathbb{P}}_{y + }}} \right)^{ - 1}}{\mathbb{R}}_{{H_x}_N}^T{{\mathbb{P}}_{x - }}{{\bf{E}}_{{z_N}}} \label{uppera}\\
&\,\,\,\,\,\,\,\,\,\,\,\,+ {\chi _1}{\left( {{{\mathbb{P}}_{x - }} \otimes {{\mathbb{P}}_{y + }}} \right)^{ - 1}}{\mathbb{R}}_{{H_x}_S}^T{{\mathbb{P}}_{x - }}\underbrace {\left( {{{\bf{E}}_{{z_S}}} - {{\mathbb{T}}_1}{{{\bf{\widehat E}}}_{{z_N}}}} \right)}_{{{{\rm{SAT\,for\,}}}}{{\bf{E}}_{{z_S}}} = {{{\bf{\widehat E}}}_{{z_N}}}}, \notag\\
&\, \frac{{d{{\bf{H}}_y}}}{{dt}} - \frac{1}{\mu } \left( {{{\mathbb{D}}_{x + }} \otimes {{\mathbb{I}}_{y - }}} \right){{\bf{E}}_z} = \label{upperb}\\
&\,\,\,\,\,\,\,\,\,\,\,\,\,\,\,\,{\sigma _{{e_1}}}{\left( {{{\mathbb{P}}_{x + }} \otimes {{\mathbb{P}}_{y - }}} \right)^{ - 1}}{\mathbb{R}}_{H{y_E}}^T{{\mathbb{P}}_{y - }}{{\bf{E}}_{{z_{E}}}} \notag\\
&\,\,\,\,\,\,\,\,\,\, + {\sigma _{{w_1}}}{\left( {{{\mathbb{P}}_{x + }} \otimes {{\mathbb{P}}_{y - }}} \right)^{ - 1}}{\mathbb{R}}_{H{y_W}}^T{{\mathbb{P}}_{y - }}{{\bf{E}}_{{z_W}}} ,\notag\\
&\frac{{d{{\bf{E}}_z}}}{{dt}}   - \frac{1}{\varepsilon } \left( {{{\mathbb{D}}_{x - }} \otimes {{\mathbb{I}}_{y - }}} \right){{\bf{H}}_y} + \frac{1}{\varepsilon } \left( {{{\mathbb{I}}_{x - }} \otimes {{\mathbb{D}}_{y - }}} \right){{\bf{H}}_x} =\label{upperc}\\
&\,\,\,\,\,\,{\chi _2}{\left( {{{\mathbb{P}}_{x - }} \otimes {{\mathbb{P}}_{y - }}} \right)^{ - 1}}{{\mathbb{\mathbb{R}}}_{{E_z}_S}}^T{{\mathbb{P}}_{x - }}\underbrace {\left( {{{\bf{H}}_{{x_S}}} - {{\mathbb{T}}_2}{{{\bf{\widehat H}}}_{{x_N}}}} \right)}_{{\rm{SAT\,for\,}}{{\bf{H}}_{{x_S}}} = {{{\bf{\widehat H}}}_{{x_N}}}},\notag
\end{align}
\end{subequations}
where ${{\mathbb{T}}_1}$ and ${{\mathbb{T}}_2}$ are interpolation matrices with dimensions of ${N_{x - }} \times {{\widehat N}_{x - }}$, which interpolate ${{{\bf{\widehat E}}}_{{z_N}}}$ and ${{{\bf{\widehat H}}}_{{x_N}}}$ from the north boundary of the bottom domain to the south boundary of the top domain. It should be noted that ${{\mathbb{T}}_1}$ and ${{\mathbb{T}}_2}$ are the same in our configuration since ${{{\bf{\widehat E}}}_{{z_N}}}$ and ${{{\bf{\widehat H}}}_{{x_N}}}$ are collocated on the north boundary of the bottom domain. However, here we use different matrices for the general purpose. ${\chi _1}$ and ${\chi _2}$ are free parameters to guarantee the stability of the subgridding technique.

In the bottom domain, the SBP-SAT FDTD formulations are expressed as  
\begin{subequations}\label{down} 
\begin{align}
&\frac{{d{{{\bf{\widehat H}}}_x}}}{{dt}} + \frac{1}{\mu }\left( {{{{\mathbb{\widehat I}}}_{x - }} \otimes {{{\mathbb{\widehat D}}}_{y + }}} \right){{{\bf{\widehat E}}}_z} = \notag\\
&\,\,\,\,\,\,\,\,\,\,\,\,{\sigma _{s_2}}{\left( {{{{\mathbb{\widehat P}}}_{x - }} \otimes {{{\mathbb{\widehat P}}}_{y + }}} \right)^{ - 1}}{\mathbb{\widehat R}}_{{H_x}_S}^T{{{\mathbb{\widehat P}}}_{x - }}{{{\bf{\widehat E}}}_{{z_S}}}\label{downa}\\
& \,\,\,\,\,\,+ {\chi _3}{\left( {{{{\mathbb{\widehat P}}}_{x - }} \otimes {{{\mathbb{\widehat P}}}_{y + }}} \right)^{ - 1}}{\mathbb{\widehat R}}_{{H_x}_N}^T{{{\mathbb{\widehat P}}}_{x - }}\underbrace {\left( {{{{\bf{\widehat E}}}_{{z_N}}} - {{\mathbb{T}}_3}{{\bf{E}}_{{z_S}}}} \right)}_{{\rm{SAT\,for\,}}{{{\bf{\widehat E}}}_{{z_N}}} = {{\bf{E}}_{{z_S}}}},\notag\\
&\frac{{d{{{\bf{\widehat H}}}_y}}}{{dt}} -\frac{1}{\mu } \left( {{{{\mathbb{\widehat D}}}_{x + }} \otimes {{{\mathbb{\widehat I}}}_{y - }}} \right){{{\bf{\widehat E}}}_z} = \notag\\
&\,\,\,\,\,\,\,\,\,\,\,\,{\sigma _{{e_2}}}{\left( {{{{\mathbb{\widehat P}}}_{x + }} \otimes {{{\mathbb{\widehat P}}}_{y - }}} \right)^{ - 1}}{\mathbb{\widehat R}}_{{H_y}_E}^T{{{\mathbb{\widehat P}}}_{y - }}{{{\bf{\widehat E}}}_{{z_E}}}\label{downb}\\
& \,\,\,\,\,\,+ {\sigma _{{w_2}}}{\left( {{{{\mathbb{\widehat P}}}_{x + }} \otimes {{{\mathbb{\widehat P}}}_{y - }}} \right)^{ - 1}}{\mathbb{\widehat R}}_{{H_y}_W}^T{{{\mathbb{\widehat P}}}_{y - }}{{{\bf{\widehat E}}}_{{z_W}}}\notag,\\
&\frac{{d{{{\bf{\widehat E}}}_z}}}{{dt}}  - \frac{1}{\varepsilon } \left( {{{{\mathbb{\widehat D}}}_{x - }} \otimes {{{\mathbb{\widehat I}}}_{y - }}} \right){{\bf{\widehat H}}_y} + \frac{1}{\varepsilon } \left( {{{{\mathbb{\widehat I}}}_{x - }} \otimes {{{\mathbb{\widehat D}}}_{y - }}} \right){{\bf{\widehat H}}_x} =  \label{downc}\\
&\,\,\,\,\,\,{\chi _4}{\left( {{{{\mathbb{\widehat P}}}_{x - }} \otimes {{{\mathbb{\widehat P}}}_{y - }}} \right)^{ - 1}}{\mathbb{\widehat R}}_{E{z_N}}^T{{{\mathbb{\widehat P}}}_{x - }}\underbrace {\left( {{{{\bf{\widehat H}}}_{{x_N}}} - {{\mathbb{T}}_4}{{\bf{H}}_{{x_S}}}} \right)}_{{\rm{SAT\,for\,}}{{{\bf{\widehat H}}}_{{x_N}}} = {{\bf{H}}_{{x_S}}}}, \notag
\end{align}
\end{subequations}
where ${{\mathbb{T}}_3}$ and ${{\mathbb{T}}_4}$ are interpolation matrices with dimensions of ${{\widehat N}_{x - }} \times {N_{x - }}$, which interpolate ${{\bf{E}}_{{z_S}}}$ and ${{\bf{H}}_{{x_S}}}$ from the south boundary of the top domain to the north boundary of the bottom domain. ${{\mathbb{T}}_3}$ and ${{\mathbb{T}}_4}$ are also the same matrices in our configuration, and we use them for the same purpose as ${{\mathbb{T}}_1}$ and ${{\mathbb{T}}_2}$. ${\chi _3}$ and ${\chi _4}$ are free parameters.

According to (\ref{upper}) and (\ref{down}), the total energy ${\cal E}$ in the computational domain is expressed as
\begin{align} \label{energytwo} 
{{\cal E}} & = \frac{1}{2}{\bf{E}}_z^T\left( {{\mathbb{P}}_{x - }^{} \otimes {\mathbb{P}}_{y - }^{}} \right){\bf{E}}_z^{} + \frac{1}{2}{\bf{H}}_y^T\left( {{\mathbb{P}}_{x + }^{} \otimes {\mathbb{P}}_{y - }^{}} \right){\bf{H}}_y^{} \notag\\
&+ \frac{1}{2}{\bf{H}}_x^T\left( {{\mathbb{P}}_{x - }^{} \otimes {\mathbb{P}}_{y + }^{}} \right){\bf{H}}_x^{}+ \frac{1}{2}{\bf{\widehat E}}_z^T\left( {{{{\mathbb{\widehat P}}}_{x - }} \otimes {{{\mathbb{\widehat P}}}_{y - }}} \right){{{\bf{\widehat E}}}_z} \notag\\
&+ \frac{1}{2}{\bf{\widehat H}}_{_y}^T\left( {{{{\mathbb{\widehat P}}}_{x + }} \otimes {{{\mathbb{\widehat P}}}_{y - }}} \right){{{\bf{\widehat H}}}_y} + \frac{1}{2}{\bf{\widehat H}}_{_x}^T\left( {{{{\mathbb{\widehat P}}}_{x - }} \otimes {{{\mathbb{\widehat P}}}_{y + }}} \right){{{\bf{\widehat H}}}_x}.
\end{align}

By taking the derivative of (\ref{energytwo}) with respect to time, and after similar mathematical manipulations in the previous section, we obtain 
\begin{align}\label{energytwod} 
&\frac{{d{\cal E}}}{{dt}} = \left( {{\sigma _{{n_1}}} - 1} \right){\bf{H}}_{{x_N}}^T{{\mathbb{P}}_{x - }}{\bf{E}}_{{z_N}}^{}\notag\\
& + \left( {{\sigma _{{w_1}}} + 1} \right){\bf{H}}_{{y_W}}^T{{\mathbb{P}}_{y - }}{\bf{E}}_{{z_W}}^{}\notag\\
& + \left( {{\sigma _{{e_1}}} - 1} \right){\bf{H}}_{{y_E}}^T{{\mathbb{P}}_{y - }}{\bf{E}}_{{z_E}}^{}{\rm{ + }}\left( {{\sigma _{{s_2}}} + 1} \right){\bf{\widehat H}}_{_{{x_S}}}^T{{{\mathbb{\widehat P}}}_{x - }}{{{\bf{\widehat E}}}_{{z_S}}}\notag\\
&+ \left( {{\sigma _{{w_2}}} + 1} \right){\bf{\widehat H}}_{_{{y_{_W}}}}^T{{{\mathbb{\widehat P}}}_{y - }}{{{\bf{\widehat E}}}_{{z_W}}} + \left( {{\sigma _{{e_2}}} - 1} \right){\bf{\widehat H}}_{_{{y_{_E}}}}^T{{{\mathbb{\widehat P}}}_{y - }}{{{\bf{\widehat E}}}_{{z_E}}}\notag\\
& + \left( {{\chi _4} + {\chi _3} - 1} \right){\bf{\widehat H}}_{_{{x_{_N}}}}^T{{{\mathbb{\widehat P}}}_{x - }}{{{\bf{\widehat E}}}_{{z_N}}}\\
& + \left( {{\chi _2} + {\chi _1} + 1} \right){\bf{H}}_{{x_S}}^T{{\mathbb{P}}_{x - }}{\bf{E}}_{{z_S}}^{}\notag\\
&- {\bf{\widehat H}}_{_{{x_{_N}}}}^T\left( {{\chi _2}{\mathbb{T}}_2^T{{\mathbb{P}}_{x - }} + {\chi _3}{{{\mathbb{\widehat P}}}_{x - }}{{\mathbb{T}}_3}} \right){\bf{E}}_{{z_S}}^{}\notag \\
& - {\bf{H}}_{{x_S}}^T\left( {{\chi _1}{{\mathbb{P}}_{x - }}{{\mathbb{T}}_1} + {\chi _4}{\mathbb{T}}_4^T{{{\mathbb{\widehat P}}}_{x - }}} \right){{{\bf{\widehat E}}}_{{z_N}}}. \notag
\end{align}

To ensure that no energy dissipation occurs in the computational domain, $d{\cal E}/dt = 0$ should be satisfied. The sufficient condition for coefficients is that ${\sigma _{n_1}}{\rm{ = }}1$, ${\sigma _{w_1}}{\rm{ =  - }}1$, ${\sigma _{e_1}}{\rm{ =  }}1$, ${\sigma _{s_2}}{\rm{ =  - }}1$, ${\sigma _{w_2}}{\rm{ =  - }}1$, ${\sigma _{e_2}}{\rm{ =  }}1$, ${\chi _2} + {\chi _1} =  - 1$, ${\chi _4} + {\chi _3}{\rm{ = }}1$. Therefore, one option is that ${\chi _1} = {\rm{ - }}1/2$, ${\chi _2} = {\rm{ - }}1/2$, ${\chi _3}{\rm{ = }}1/2$, and ${\chi _4}{\rm{ = }}1/2$. These values are used in the simulations in Section V. Under these conditions, we can have additional conditions for interpolation matrices, which can be expressed as 
\begin{subequations}\label{Tequ}
\begin{align}
&{{\mathbb{T}}_3} = {\left( {{{{\mathbb{\widehat P}}}_{x - }}} \right)^{ - 1}}{\mathbb{T}}_2^T{{\mathbb{P}}_{x - }},\label{Tequa}\\ 
&{\mathbb{T}}_4^T = {{\mathbb{P}}_{x - }}{{\mathbb{T}}_1}{\left( {{{{\mathbb{\widehat P}}}_{x - }}} \right)^{ - 1}} . \label{Tequb} 
\end{align}
\end{subequations}

(\ref{Tequ})  are norm compatible conditions, which can be found similar relationship in \cite{HOFDF} \cite{SAAIO}, and the interpolation matrices should be satisfied to guarantee the stability of the proposed subgridding scheme.   

\subsection{Derivation of Interpolation Matrices}
In this subsection, the interpolation matrices in the SBP-SAT FDTD subgridding scheme is derived. Since four interpolation matrices (actually only two since ${{\mathbb{T}}_1} = {{\mathbb{T}}_2}$ and ${{\mathbb{T}}_3} = {{\mathbb{T}}_4}$) are required to be determined and they are related through  (\ref{Tequ}), only two of them, e.g., ${\mathbb{T}}_1$ and ${\mathbb{T}}_2$, are determined, and other two matrices can be calculated from (\ref{Tequ}). Assume that the fine grids are applied in the bottom domain, which implies that ${\mathbb{T}}_1$ and ${\mathbb{T}}_2$ are the interpolation matrices from fine to coarse meshes, ${\mathbb{T}}_3$ and ${\mathbb{T}}_4$ are the interpolation matrices from coarse to fine meshes. ${\mathbb{T}}_1$ can be expressed as (\ref{IF2C}) at the bottom of this page according to \cite{SAAIO}. 
\begin{align}\label{IF2C}
{{\mathbb{T}}_1} = \left[ {\begin{array}{*{20}{c}}
	{{a_{1,1}}}&{{a_{1,2}}}&{{a_{1,3}}}&{{a_{1,4}}}&{{a_{1,5}}}&{{a_{1,6}}}&{}&{}&{}&{}&{}&{}&{}&{}&{}&{}&{}&{}\\
	{{a_{2,1}}}&{{a_{2,2}}}&{{a_{2,3}}}&{{a_{2,4}}}&{{a_{2,5}}}&{{a_{2,6}}}&{}&{}&{}&{}&{}&{}&{}&{}&{}&{}&{}&{}\\
	{{a_{3,1}}}&{{a_{3,2}}}&{{a_{3,3}}}&{{a_{3,4}}}&{{a_{3,5}}}&{{a_{3,6}}}&{}&{}&{}&{}&{}&{}&{}&{}&{}&{}&{}&{}\\
	{}&{}&{}&{}&{{a_2}}&{{a_1}}&{{a_1}}&{{a_2}}&{}&{}&{}&{}&{}&{}&{}&{}&{}&{}\\
	{}&{}&{}&{}&{}&{}&{{a_2}}&{{a_1}}&{{a_1}}&{{a_2}}&{}&{}&{}&{}&{}&{}&{}&{}\\
	{}&{}&{}&{}&{}&{}&{}&{}&{...}&{...}&{...}&{...}&{}&{}&{}&{}&{}&{}\\
	{}&{}&{}&{}&{}&{}&{}&{}&{}&{}&{{a_2}}&{{a_1}}&{{a_1}}&{{a_2}}&{}&{}&{}&{}\\
	{}&{}&{}&{}&{}&{}&{}&{}&{}&{}&{}&{}&{{a_{3,6}}}&{{a_{3,5}}}&{{a_{3,4}}}&{{a_{3,3}}}&{{a_{3,2}}}&{{a_{3,1}}}\\
	{}&{}&{}&{}&{}&{}&{}&{}&{}&{}&{}&{}&{{a_{2,6}}}&{{a_{2,5}}}&{{a_{2,4}}}&{{a_{2,3}}}&{{a_{2,2}}}&{{a_{2,1}}}\\
	{}&{}&{}&{}&{}&{}&{}&{}&{}&{}&{}&{}&{{a_{1,6}}}&{{a_{1,5}}}&{{a_{1,4}}}&{{a_{1,3}}}&{{a_{1,2}}}&{{a_{1,1}}}
	\end{array}} \right] 
\end{align}
It should be noted that the second-order accuracy is for inner nodes and the first-order accuracy is for boundary closure nodes.  

To calculate the interpolation matrices, an optimal $L_2$ error is introduced as
\setcounter{equation}{16}
\begin{align}\label{error}
{\bf{e}}_c^k = {{\mathbb{T}}_1}{\bf{x}}_f^{\rm{k}} - {\bf{x}}_c^{\rm{k}},\,\,\,\,{\mkern 1mu} {\mkern 1mu} {\mkern 1mu} {\bf{e}}_f^k = {{\mathbb{T}}_3}{\bf{x}}_c^{\rm{k}} - {\bf{x}}_f^{\rm{k}},\,\,\,\,k= 0,1,
\end{align}
where ${\bf x}_f$ and ${\bf x}_c$ refer to the coordinates of the nodes near the boundaries in the fine and coarse grids, respectively. The $L_2$ error should vanish for $k = 0, 1$ in the interior region and for $k = 0, 1$ near the boundary. Therefore, the accuracy of interpolation matrices keep consistency with the SBP operators \cite{SAAIO}. If there are still free parameters to be determined, they can be calculated by minimizing 
\begin{align}\label{minierror}
{e_k^2 = {\left( {{\bf{e}}_c^k} \right)^T} \cdot {\bf{e}}_c^k + {\left( {{\bf{e}}_f^k} \right)^T} \cdot {\bf{e}}_f^k,\,\,\,\, k \ge 2}. 
\end{align}

Through the optimization algorithms, such as the genetic algorithm (GA) \cite{GA}, the particle swarm optimization (PSO) \cite{PSO}, until all free parameters are solved. The flowchart to calculate interpolation matrices mentioned in this section is shown in Fig. \ref{flowchart}. In the appendix, we listed ${\mathbb{T}}_1$ used in our simulations, and other three matrices can be easily obtained through (\ref{Tequ}).

\begin{figure} 	
	\centerline{\includegraphics[scale=0.5]{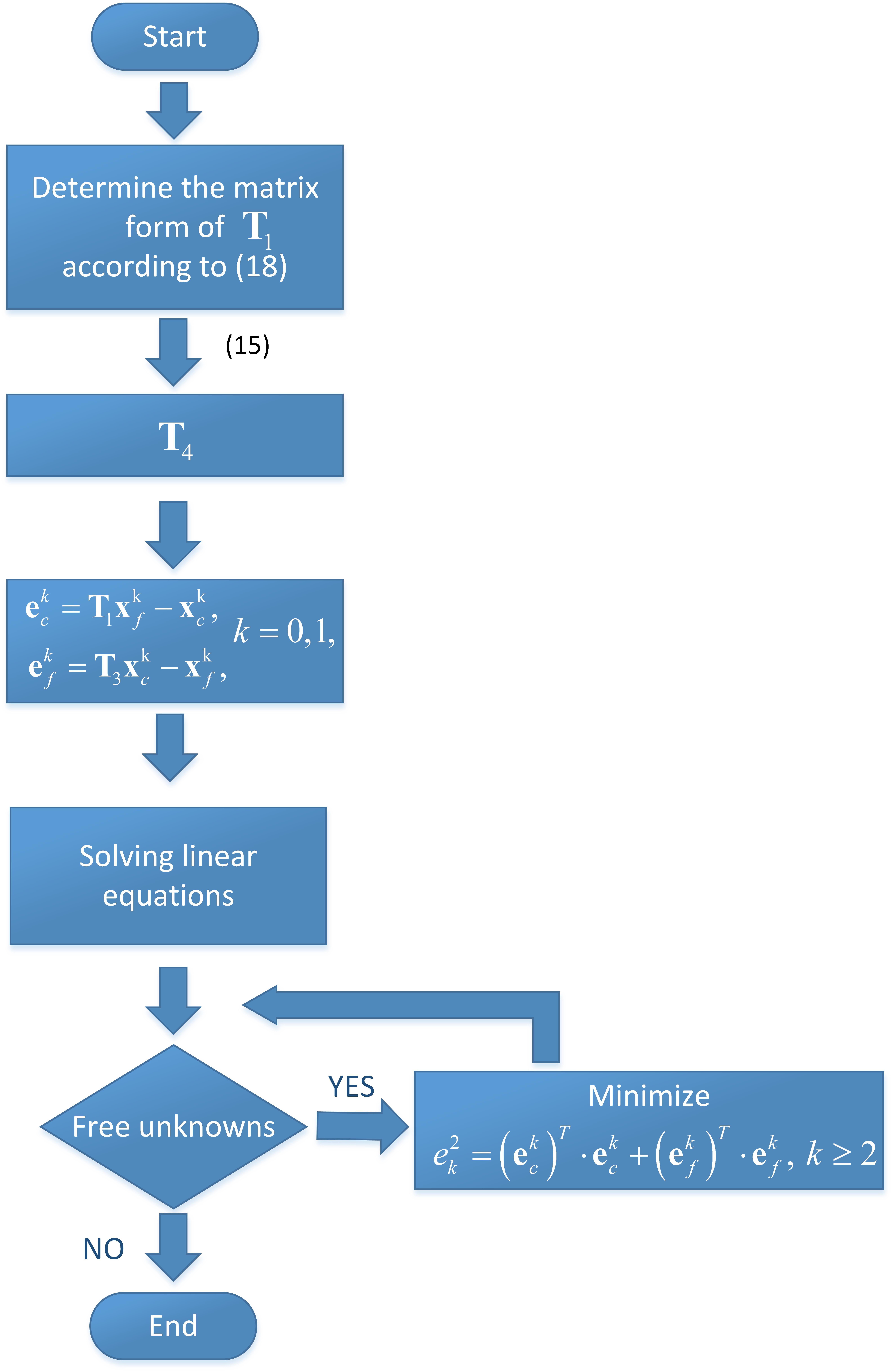}}	
	\caption{The flowchart to derive the interpolation matrices, ${\mathbb{T}_1}$ and ${\mathbb{T}_4}$.  } 
	\label{flowchart} 
\end{figure}

\subsection{Mesh Block Division in the Computational Domain}
In the practical implementations, the subgridding domains are embedded in coarse mesh domains. To make the above analysis valid, we can divide them into rectangular subdomains. Fig. 5 shows one such possibility to divide the computational domains into several subdomains, where the discrete operators satisfy the SBP property. Therefore, the stability of the proposed FDTD method can be guaranteed. The corresponding SATs and the interpolation matrices obtained from the previous subsection are used to couple the interfaces between coarse and fine mesh blocks.

\begin{figure} \label{four}	
	\centerline{\includegraphics[scale=0.5]{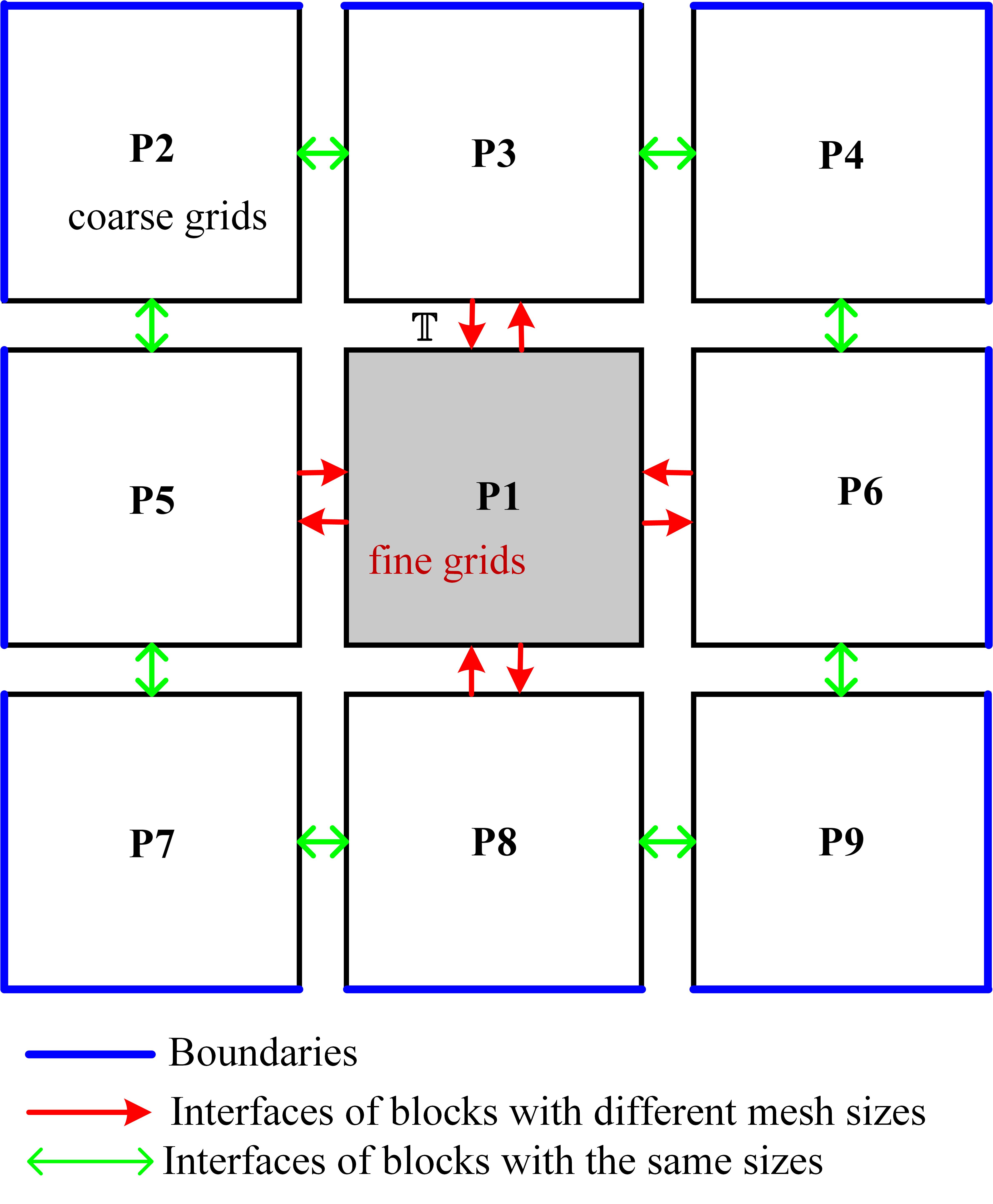}}	
	\caption{One typical domain subdivision scheme for the proposed SBP-SAT FDTD subgridding scheme. The boundary conditions between coarse and fine grids, such as P1 and P3, are handled through the interpolation matrices ${\mathbb{T}}$, and the SAT techniques are used to handle the boundary conditions between the grids with the same sizes, such as P2 and P3.  } 
\end{figure}

\section{THE CFL CONDITION FOR THE PROPOSED SBP-SAT FDTD METHOD}
Since the explicit second-order finite-difference scheme is used in the proposed SBP-SAT FDTD method, time steps are also constrained by the cell sizes and the boundary conditions like the traditional FDTD method. In this section, the stability condition of the proposed SBP-SAT FDTD method is derived. In the following derivation, the PEC boundary conditions are assumed.

According to (\ref{sbpsatequ}) and the selected values of the corresponding free parameters, the semi-discrete formulations for the SBP-SAT FDTD method with the PEC boundary conditions can be expressed as
\begin{subequations}\label{PEC}
	\begin{align} 	
	&\frac{{d{{\bf{H}}_x}}}{{dt}} +  \frac{1}{\mu }\left( {{{\mathbb{I}}_{x - }} \otimes {{\mathbb{D}}_{y + }}} \right){{\bf{E}}_z} =  - {\left( {{{\mathbb{P}}_{x - }} \otimes {{\mathbb{P}}_{y + }}} \right)^{ - 1}}{\bf{E}}_{zs}^T{{\mathbb{P}}_{x - }}{{\bf{E}}_z}_s\notag\\
	&\,\,\,\,\,\,\,\,\,\,\,\,\,\,\,\,\,\,\,\,\,\,\,\,\,\,+ {\left( {{{\mathbb{P}}_{x - }} \otimes {{\mathbb{P}}_{y + }}} \right)^{ - 1}}{\bf{E}}_{zn}^T{{\mathbb{P}}_{x - }}{{\bf{E}}_z}_n ,\label{Za}\\
	&\frac{{d{{\bf{H}}_y}}}{{dt}} -   \frac{1}{\mu }\left( {{{\mathbb{D}}_{x + }} \otimes {{\mathbb{I}}_{y - }}} \right){{\bf{E}}_z} = {\left( {{{\mathbb{P}}_{x + }} \otimes {{\mathbb{P}}_{y - }}} \right)^{ - 1}}{\bf{E}}_{ze}^T{{\mathbb{P}}_{y - }}{{\bf{E}}_z}_e\notag\\
	&\,\,\,\,\,\,\,\,\,\,\,\,\,\,\,\,\,\,\,\,\,\,\,\,\,\,- {\left( {{{\mathbb{P}}_{x + }} \otimes {{\mathbb{P}}_{y - }}} \right)^{ - 1}}{\bf{E}}_{zw}^T{{\mathbb{P}}_{y - }}{{\bf{E}}_z}_w.\label{Zb}\\
	&\frac{{d{{\bf{E}}_z}}}{{dt}} - \frac{1}{\varepsilon }\left( {{{\mathbb{D}}_{x - }} \otimes {{\mathbb{I}}_{y - }}} \right){{\bf{H}}_y} + \frac{1}{\varepsilon }\left( {{{\mathbb{I}}_{x - }} \otimes {{\mathbb{D}}_{y - }}} \right){{\bf{H}}_x} = {\bf{0}}, \label{Zc} 
	\end{align}
\end{subequations}

According to the second-order centre-difference scheme  
\begin{align} \label{Eneru}
\frac{{d{{\bf{U}}^n}}}{{dt}}{\rm{ = }}\frac{{{{\bf{U}}^{n + \frac{1}{2}}} - {{\bf{U}}^{n - \frac{1}{2}}}}}{{\Delta t}},
\end{align}
the time-marching formulations of (\ref{PEC}) are rewritten as 
\begin{subequations}\label{TIMEMARCH}
\begin{align} 
{\bf{H}}_x^{n + \frac{3}{2}} &= {\bf{H}}_x^{n + \frac{1}{2}} - \frac{{\Delta t}}{\mu }\left( {{{\mathbb{I}}_{x - }} \otimes {{\mathbb{D}}_{y + }}} \right){\bf{E}}_z^{n + 1}\notag\\
&- \frac{{\Delta t}}{\mu }{\left( {{{\mathbb{P}}_{x - }} \otimes {{\mathbb{P}}_{y + }}} \right)^{ - 1}}{\bf{E}}_{zs}^T{{\mathbb{P}}_{x - }}{{\mathbb{A}}_s}{\bf{E}}_z^{n + 1} \label{TIMEMARCHa}\\
&+ \frac{{\Delta t}}{\mu }{\left( {{{\mathbb{P}}_{x - }} \otimes {{\mathbb{P}}_{y + }}} \right)^{ - 1}}{\bf{E}}_{zn}^T{{\mathbb{P}}_{x - }}{{\mathbb{A}}_n}{\bf{E}}_z^{n + 1},\notag\\
{\bf{H}}_y^{n + \frac{3}{2}} &= {\bf{H}}_y^{n + \frac{1}{2}} + \frac{{\Delta t}}{\mu }\left( {{{\mathbb{D}}_{x + }} \otimes {{\mathbb{I}}_{y - }}} \right){\bf{E}}_z^{n + 1}\notag\\
&- \frac{{\Delta t}}{\mu }{\left( {{{\mathbb{P}}_{x + }} \otimes {{\mathbb{P}}_{y - }}} \right)^{ - 1}}{\bf{E}}_{zw}^T{{\mathbb{P}}_{y - }}{{\mathbb{A}}_w}{\bf{E}}_z^{n + 1} \label{TIMEMARCHb}\\
&+ \frac{{\Delta t}}{\mu }{\left( {{{\mathbb{P}}_{x + }} \otimes {{\mathbb{P}}_{y - }}} \right)^{ - 1}}{\bf{E}}_{ze}^T{{\mathbb{P}}_{y - }}{{\mathbb{A}}_e}{\bf{E}}_z^{n + 1}\notag,\\
{\bf{E}}_z^{n + 1} &= {\bf{E}}_z^n + \frac{{\Delta t}}{\varepsilon }\left( {{{\mathbb{D}}_{x - }} \otimes {{\mathbb{I}}_{y - }}} \right){\bf{H}}_y^{n + \frac{1}{2}} \notag\\
&- \frac{{\Delta t}}{\varepsilon }\left( {{{\mathbb{I}}_{x - }} \otimes {{\mathbb{D}}_{y - }}} \right){\bf{H}}_x^{n + \frac{1}{2}} ,\label{TIMEMARCHc}
\end{align}
\end{subequations} 
where $\Delta t$ is the time step, $n$ is the $n$th time step in the simulation.
 
To make the following derivation clear, we rewrite (\ref{TIMEMARCH}) into the matrix form
\begin{subequations} \label{MATRIXFORM}
\begin{align} 
 {\bf{H}}^{n + \frac{3}{2}} &= {\bf{H}}_{}^{n{\rm{ + }} \frac{1}{2}} + \Delta t{{\mathbb{D}}_E}{\bf{E}}_{}^{n + 1},\label{a}\\
 {\bf{E}}^{n + 1} &= {\bf{E}}_{}^n - \Delta t{{\mathbb{D}}_H}{\bf{H}}_{}^{n + \frac{1}{2}},\label{b}
\end{align}
\end{subequations} 
where ${\bf{H}} = {\left[ {\begin{array}{*{20}{c}}{{{\bf{H}}_x}},&{{{\bf{H}}_y}}
\end{array}} \right]^T}$, 
\begin{align}	
{{\mathbb{D}}_H}{\rm{ = }}\frac{1}{\varepsilon }\left[ {\begin{array}{*{20}{c}}{\left( {{{\mathbb{I}}_{x - }} \otimes {{\mathbb{D}}_{y - }}} \right)}&{ - \left( {{{\mathbb{D}}_{x - }} \otimes {{\mathbb{I}}_{y - }}} \right)}\end{array}} \right],\notag
\end{align}
\begin{align}
 {{\mathbb{D}}_E}{\rm{ = }}\frac{1}{\mu }\left[ {\begin{array}{*{20}{c}}
	{\left[ \begin{array}{l}
		- \left( {{{\mathbb{I}}_{x - }} \otimes {{\mathbb{D}}_{y + }}} \right) \notag\\
		- {\left( {{{\mathbb{P}}_{x - }} \otimes {{\mathbb{P}}_{y + }}} \right)^{ - 1}}{\bf{E}}_{zs}^T{{\mathbb{P}}_{x - }}{{\mathbb{A}}_s}\notag\\
		+ {\left( {{{\mathbb{P}}_{x - }} \otimes {{\mathbb{P}}_{y + }}} \right)^{ - 1}}{\bf{E}}_{zn}^T{{\mathbb{P}}_{x - }}{{\mathbb{A}}_n}
		\end{array} \right]}\notag\\
	{\left[ \begin{array}{l}
		\left( {{{\mathbb{D}}_{x + }} \otimes {{\mathbb{I}}_{y - }}} \right) \notag\\
		- {\left( {{{\mathbb{P}}_{x + }} \otimes {{\mathbb{P}}_{y - }}} \right)^{ - 1}}{\bf{E}}_{zw}^T{{\mathbb{P}}_{y - }}{{\mathbb{A}}_w}\notag\\
		+ {\left( {{{\mathbb{P}}_{x + }} \otimes {{\mathbb{P}}_{y - }}} \right)^{ - 1}}{\bf{E}}_{ze}^T{{\mathbb{P}}_{y - }}{{\mathbb{A}}_e}
		\end{array} \right]}
	\end{array}} \right] \notag.
\end{align}
By rewriting (\ref{MATRIXFORM}) into a single matrix equation, we obtain
 \begin{align}  \label{G}
 {{\bf{U}}^{n{\rm{ + }}1}}{\rm{ = }}{\mathbb{G}}{{\bf{U}}^n},
 \end{align}
 where  
 \begin{align} \label{GU}
  {\mathbb{G}}{\rm{ = }}\left[ {\begin{array}{*{20}{c}}
 	{\mathbb{I}}&{ - \Delta t{{\mathbb{D}}_H}}\\
 	{\Delta t{{\mathbb{D}}_E}}&{{\mathbb{I}} - \Delta {t^2}{\mathbb{M}}}
 	\end{array}} \right] ,{\mathbb{M}}{\rm{ = }}{{\mathbb{D}}_H}{{\mathbb{D}}_E},
 {{\bf{U}}^n} = \left[ {\begin{array}{*{20}{c}}
 	{{\bf{E}}_{}^n}\\
 	{{\bf{H}}_{}^{n{\rm{ + }}1/2}}
 	\end{array}} \right]. 
 \end{align}
 
According to the analysis of the stability in the FDTD method in \cite{CFLC}, the time step in the SBP-SAT FDTD method should be under the following condition  
 \begin{align} \label{CFLC}   
  \Delta t \le \frac{2}{{\sqrt {{\lambda _{\max }}\left( {\mathbb{M}} \right)} }},  
\end{align}   
where ${\lambda _{\max }}\left( {\mathbb{A}} \right)$ indicates the maximum eigenvalue of $\mathbb{A}$.

\section{NUMERICAL RESULTS AND DISCUSSION}
In this section, three numerical examples including a PEC cavity, an iris filter, and the specific absorption rate (SAR) calculation for the human head are carried out to validate the stability, accuracy and efficiency of the proposed SBP-SAT FDTD method and the proposed FDTD subgridding method. All codes are written in Matlab and run on a workstation with an Intel i7-7700 3.6 GHz CPU and 256 G memory. A single thread is used in our simulations for fair comparison.

\subsection{A Two-dimensional Cavity with PEC Boundaries}   
\begin{figure}  
	\centerline{\includegraphics[scale=0.5]{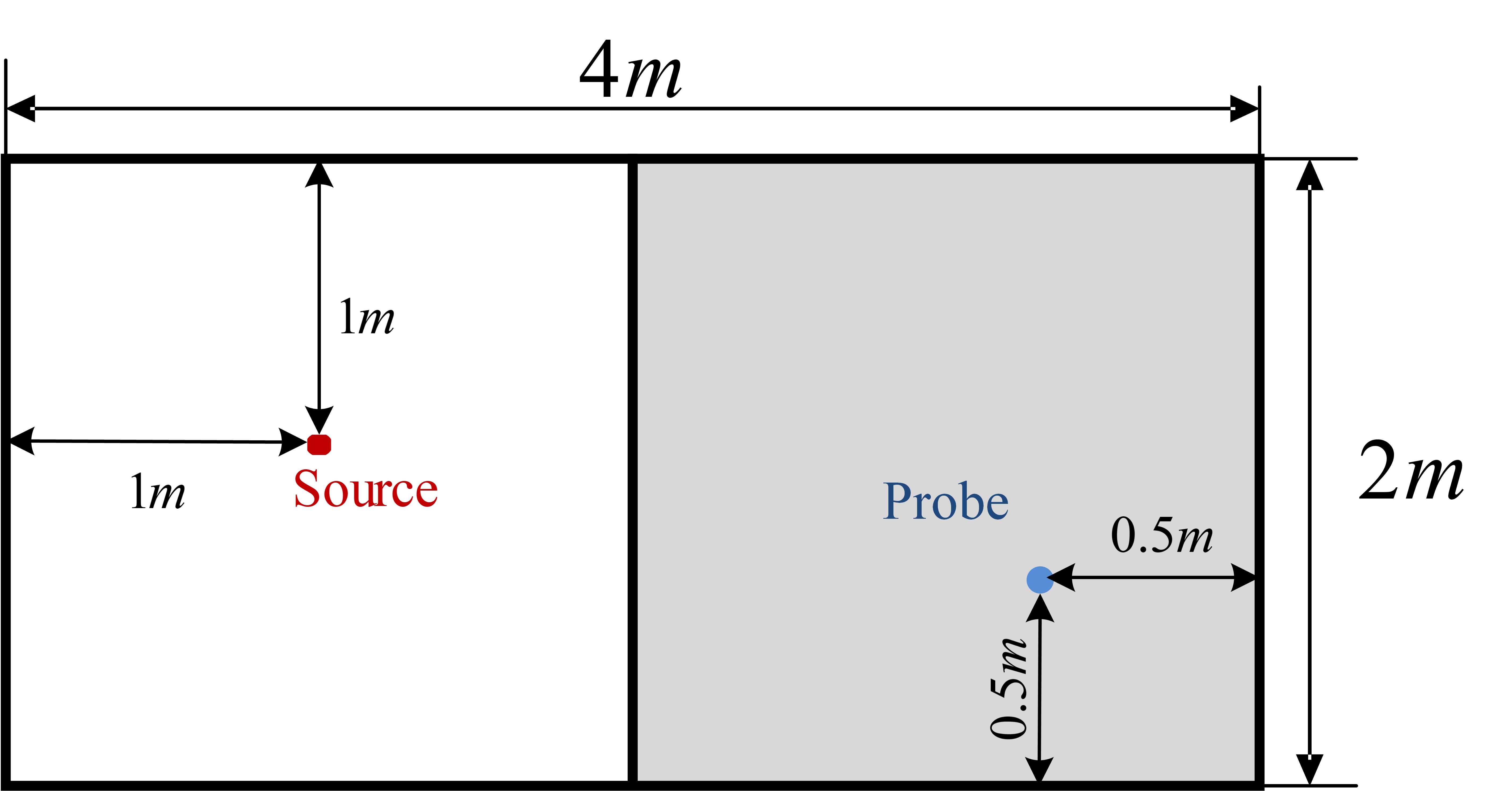}}	
	\caption{Geometrical configuration of the cavity used in the proposed subgridding method, and the source and probe locations are placed at (1,1)[m], and (3,0.5)[m], respectively.}
	\label{CAVITYGEO}	
\end{figure}

A two-dimensional cavity with the PEC boundaries is first considered. The size of the cavity is $4$ m in length and $2$ m in width, as shown in Fig. \ref{CAVITYGEO}. It is filled with the air. A Gaussian pulse with the bandwidth of $1$ GHz is used as the excitation source, which is located at $(2,2)$[m]. A probe is placed at $(3,0.5)$[m] to record the transient electric fields in the simulations. 

Two scenarios are considered in this numerical example. One is that the cavity is discretized with the uniform meshes, with the mesh size $\Delta {\rm{ = }}4 \times {10^{{\rm{ - }}2}}$ m in both $x$ and $y$ direction. The other is that the cavity is divided into two domains, as shown in Fig. \ref{CAVITYGEO}. The left domain is discretized with the uniform meshes with the mesh size $\Delta {\rm{ = }}4 \times {10^{{\rm{ - }}2}}$ m, and the uniform meshes with the mesh size $\Delta {\rm{ = }}2 \times {10^{{\rm{ - }}2}}$ m are used in the right domain. The time steps used in the simulations are $0.99$ times of the CFL condition in the fine mesh domain defined in (\ref{CFLC}).  

\begin{figure}  
  	\begin{minipage} {0.48\linewidth}\label{sixb}
  		\centerline{\includegraphics[scale=0.205]{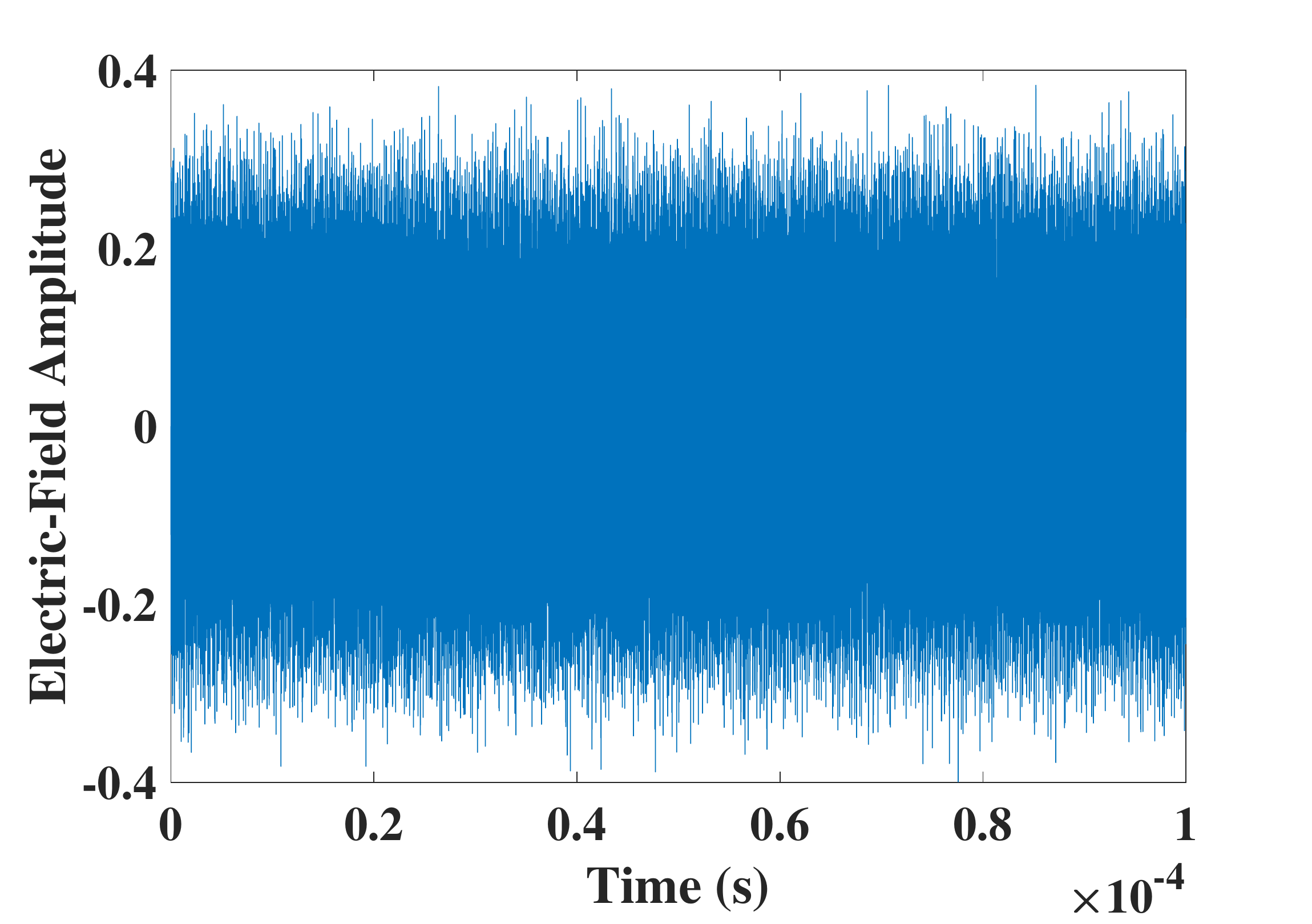}}
  		\centerline{(a)}
  	\end{minipage}
  	\centering  	
  \begin{minipage} {0.48\linewidth}\label{six}
  	\centerline{\includegraphics[scale=0.205]{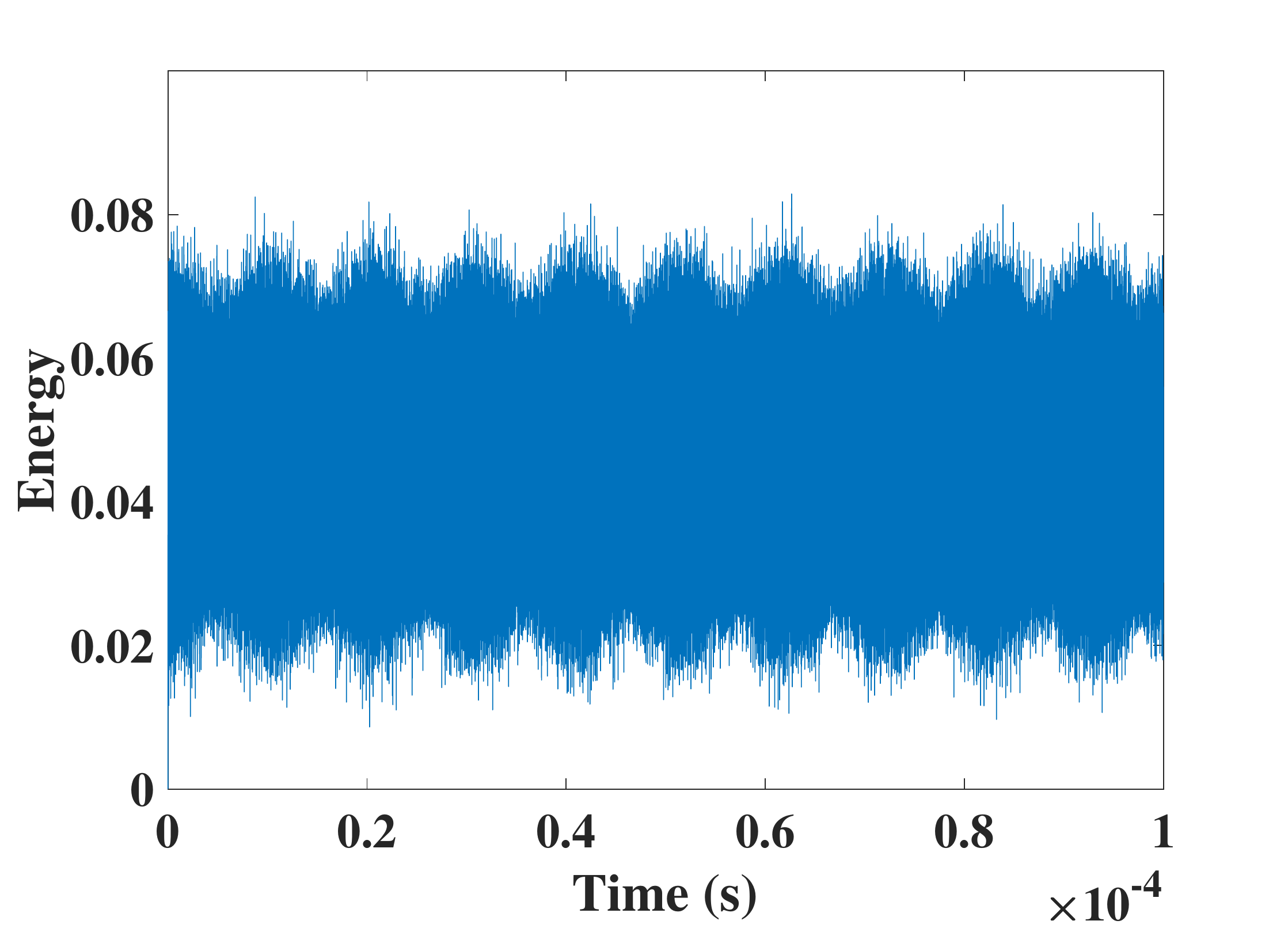}}
  	\centerline{(b)}
  \end{minipage}
  	\centering
  	\caption{(a) The electric fields and (b) the energy verse the time at the probe obtained from the SBP-FDTD method.}
  	\label{ONEDOMAIN}
  \end{figure}

To investigate the stability of the proposed SBP-SAT FDTD method and the subgridding SBP-SAT FDTD method, we calculated the electric fields and the electromagnetic energy defined in (\ref{energysbp}) in the whole time duration. The simulation time is $1.0 \times {10^{-4}}$s. Therefore, the overall counts of time steps in the two scenarios are  $1.2 \times {10^6}$. Fig. \ref{ONEDOMAIN} shows the electric fields and the energy recorded at the probe location. It is easy to find that both the electric fields and the energy obtained from the SBP-SAT FDTD method is stable after an extremely long time simulation. Since the discrete partial differential operators used in the proposed SBP-SAT FDTD method satisfy the SBP property and the SATs are carefully selected, its long-time stability is theoretically guaranteed. The numerical results agree well with our previous analysis and confirm the stability.

\begin{figure} 	
	\begin{minipage}[h]{0.48\linewidth}\label{sevenb}
		\centerline{\includegraphics[scale=0.205]{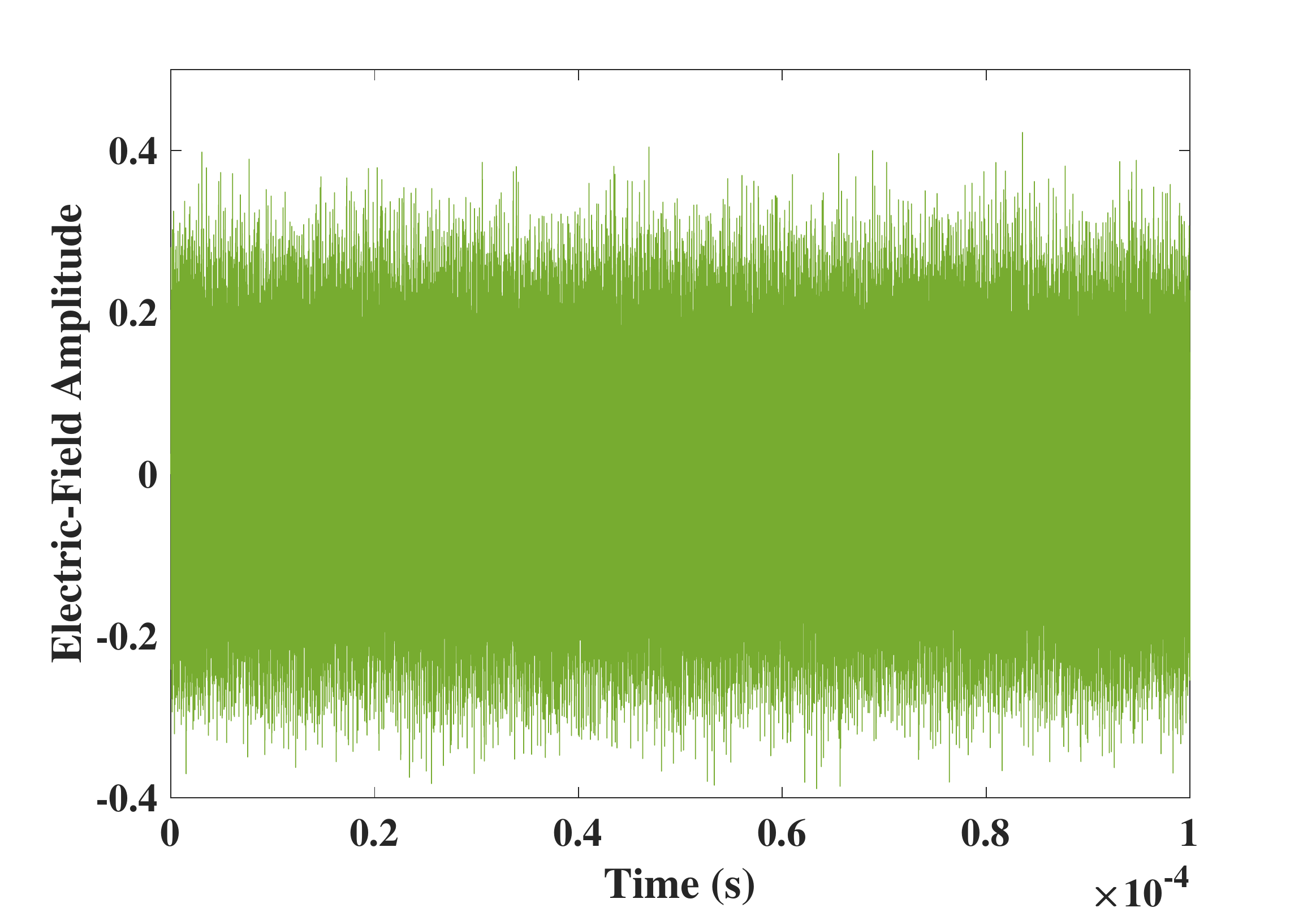}}
		\centerline{(a)}
	\end{minipage}
	\centering
	\begin{minipage}[h]{0.48\linewidth}\label{seven}
		\centerline{\includegraphics[scale=0.205]{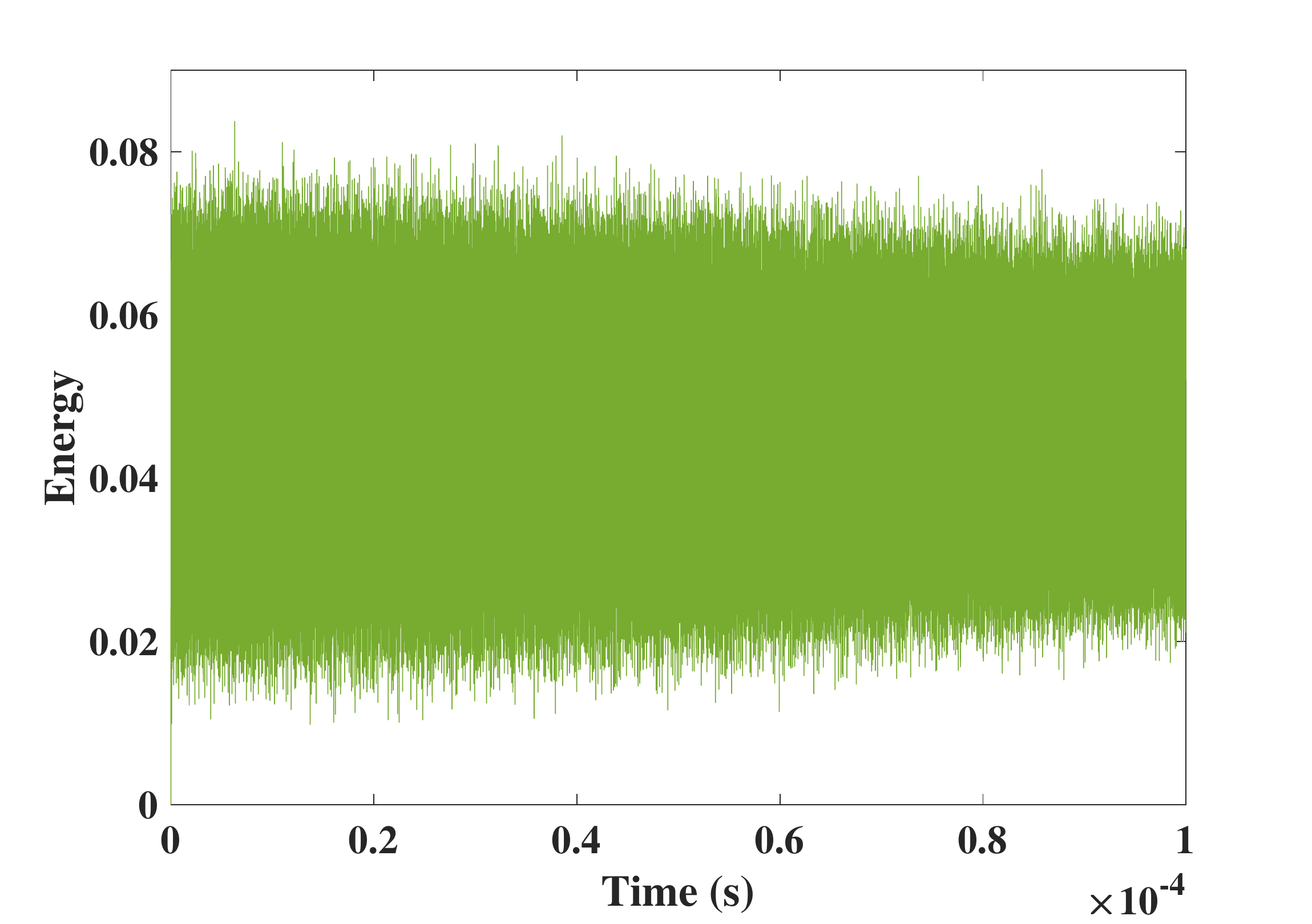}}
		\centerline{(b)}
	\end{minipage}
	\centering
	\caption{(a) The electric fields and (b) the energy verse time at the probe obtained from the proposed subgridding method. }
	\label{SUBGRIDDINGRES}
\end{figure}

Fig. \ref{SUBGRIDDINGRES}(a)-(b) shows the electric fields and the energy obtained from the proposed subgridding FDTD method. It can be found that both the electric fields and the energy are remain finite in the simulation. The SATs are used to weakly enforce the boundary conditions, ${\bf{E}}_{{z_E}} = {{{\bf{\widehat E}}}_{{z_W}}}$ and ${{\bf{H}}}_{y_E} = {{\bf{\widehat H}}_{{y_W}}}$. The free parameters and the interpolation matrices are carefully designed to guarantee the stability of the proposed subgridding FDTD method when multiple mesh blocks with different mesh sizes are used. The numerical results confirm our previous analytical analysis. An interesting observation is that the energy in the computational domain seems to gradually converge to a constant value. The reason for it may be that the fields are well resolved after the subgridding meshes are used.

 	
\begin{figure}   
	\centerline{\includegraphics[scale=0.2]{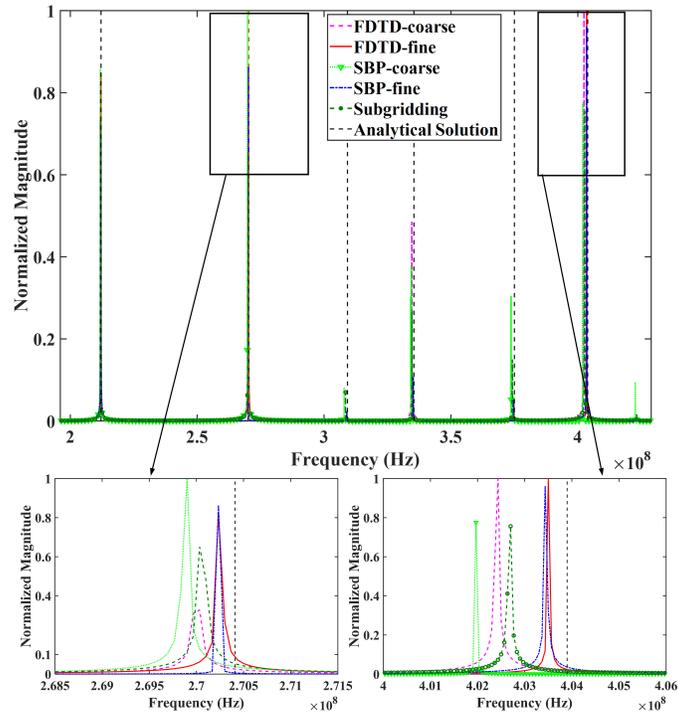}}	
	\caption{Resonant frequencies for TM modes obtained from the FDTD method, the SBP-SAT FDTD method, and the proposed FDTD subgridding method.}
	\label{eight}	
\end{figure}

The resonant frequencies of the cavity are calculated to validate the accuracy of the SBP-SAT FDTD method and the FDTD subgridding method, respectively, which are obtained through the discrete Fourier transform of the transient electric fields. The results obtained from the FDTD method with fine mesh sizes $\Delta {\rm{ = }}2 \times {10^{{\rm{ - }}2}}$ m and the analytical results are also plotted for comparison purposes. As shown in Fig. 9, the results obtained from the FDTD method and the SPB-SAT FDTD method with fine meshes show excellent agreement with the analytical resonant frequencies. However, when coarse meshes are used in the computational domain, results obtained from the two methods show slightly large discrepancies compared with the analytical ones. Since the second-order finite-difference scheme is used in the proposed SBP-SAT FDTD method and the FDTD method, they can obtain the same level of accuracy. When the subgridding meshes are used in the simulation, the accuracy of the resonant frequencies obtained from the proposed subgridding FDTD method is improved compared with those obtained from the FDTD method and the SBP-SAT FDTD method with coarse meshes.

\subsection{An IRIS Filter}    
\begin{figure}  
  \centerline{\includegraphics[scale=0.37]{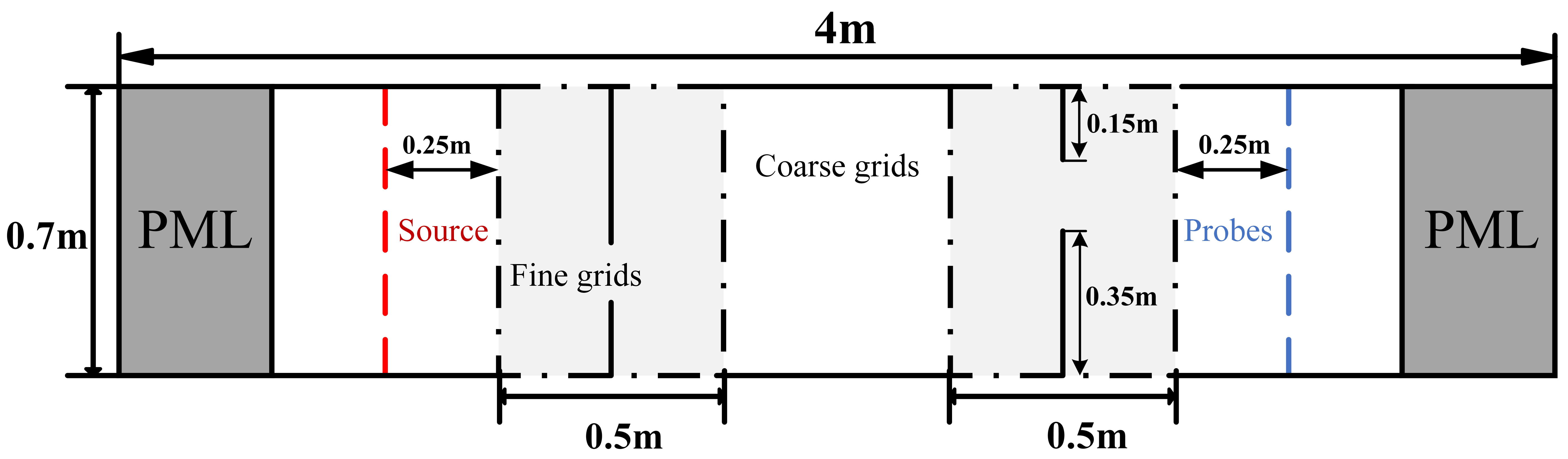}}  	
  \caption{The geometrical configuration of the iris waveguide. }
  \label{GEOIRIS}	
\end{figure}

An iris filter is considered to validate the accuracy and efficiency the proposed SBP-FDTD method and the proposed FDTD subgridding method. As shown in Fig. \ref{GEOIRIS}, the whole computational domain is $4$ m in length and $0.7$ m in width. To support the propagation modes in this iris filter, the boundary conditions and the two iris are assigned as the perfect magnetic conductor (PMC). The two iris are $0.2$ m in thickness and are placed in the two subgridding areas, which are marked in light gray domains in Fig. \ref{GEOIRIS}. Two subgridding domains are $0.5$ m in length and $0.7$ m in width. The cell sizes of coarse meshes are $\Delta x = \Delta y = 0.05$ m, and $\Delta x = \Delta y = 0.025$ m for the subgridding domains. A 10-layered perfectly matched layer (PML) \cite{CPML} is used to truncate the left and right of the computational domain. A line current source is a Gaussian pulse with $0.4$ GHz in bandwidth, which is located at $x=1$ m, and the probe is placed at $x=3.8$ m. The time steps used in the simulations are $0.99$ times of the CFL condition in the fine mesh domain.

\begin{figure} 
	\centerline{\includegraphics[scale=0.24]{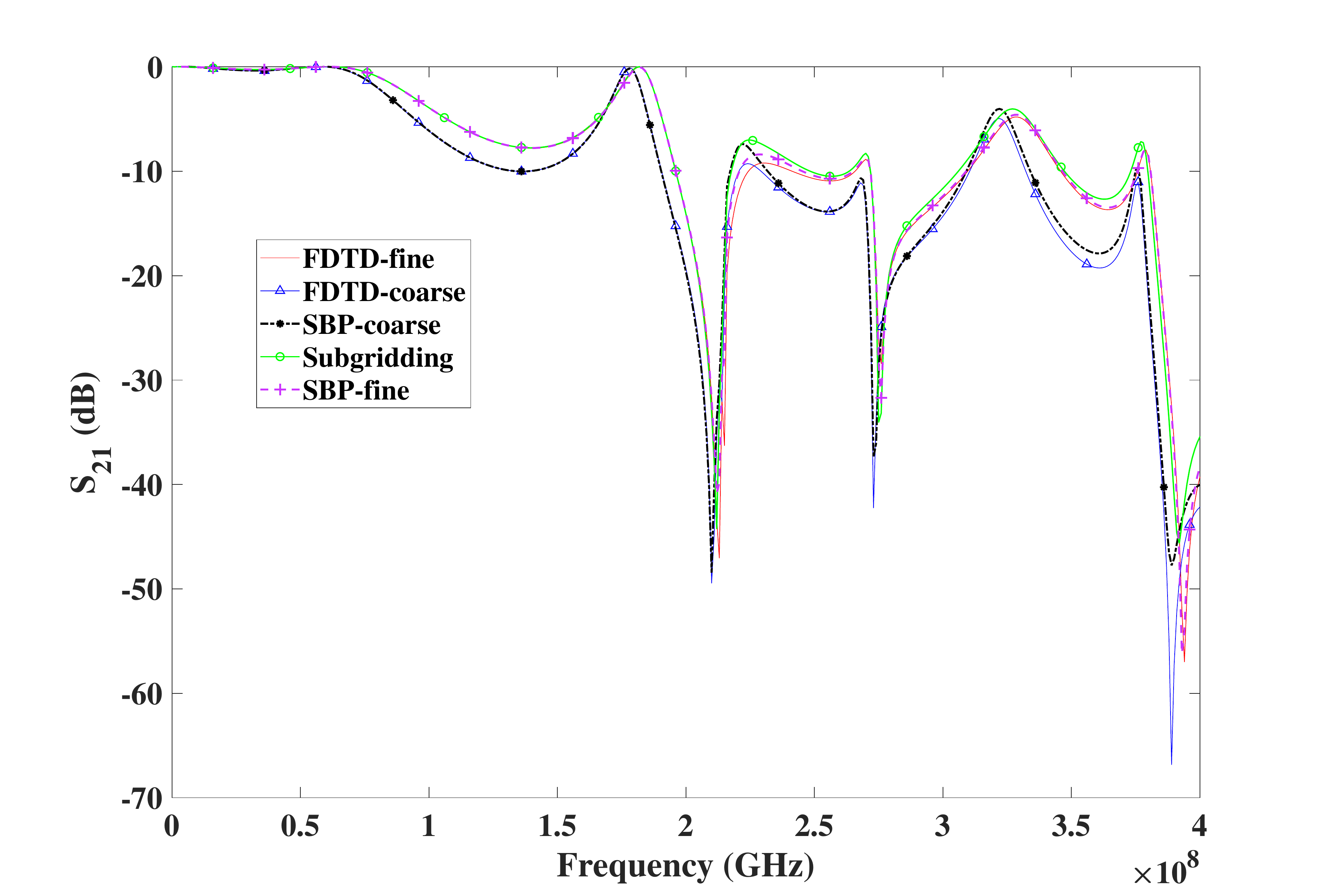}}
	\caption{The frequency response of the iris waveguide obtained from the FDTD method, the SBP-SAT FDTD method, and the SBP-SAT FDTD subgridding method.}
	\label{FRERESPONSE}	
\end{figure} 
Fig. \ref{FRERESPONSE} show the frequency responses from DC to $400$ MHz obtained from the FDTD method, the SBP-SAT FDTD method, and the proposed FDTD subgridding method. It can be found that results obtained from the FDTD method and the proposed SBP-SAT FDTD method with fine meshes show excellent agreement. However, when coarse meshes are used in the two methods, significant discrepancies exist in the frequency responses, especially in the high frequency regions, such as from $210$ MHz to $260$ MHz, and from $320$ MHz to $360$ MHz. When the same meshes are used in the two methods, the same level of accuracy can be achieved. When the subgridding meshes are used, results obtained from the proposed FDTD subgridding method show significantly accuracy improvement compared with those from the FDTD method and the SBP-SAT FDTD method with coarse meshes.   

\begin{table}
	\centering
	\caption{Detailed parameters of different tissues used in the SAR calculation}\label{parameter}
	\begin{tabular}{lccc}
		\hline
		\hline
		\textbf{Tissue} &\textbf{Density [${\bf{kg/{m^3}}}$]  }  & \multicolumn{2}{c}{\textbf{Dielectric Properties}}\cr  
		\hline
		\hline
		&\textbf{Average} &\textbf{$\varepsilon_r$}&\textbf{$\sigma$ [S/m]}\cr
		\hline
		Brain                &   1,046          & 4    &   0.04   \\
		\hline
		Cerebrospinal Fluid            &  1,007   & 4    & 2     \\
		\hline
		Dura             &   1,174        & 4      &  0.5      \\
		\hline
		Skull            &   1,908      & 2.5   &  0.02   \cr
		\hline 
		\hline 
	\end{tabular}
\end{table}

\subsection{The SAR Calculation}  
\begin{figure}
	\centerline{\includegraphics[scale=0.5]{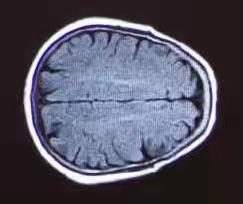}}  	
	\caption{The CT image of the cross-sectional human head used in the simulation.}
	\label{HUMANCT}	
\end{figure}

\begin{figure}
  	\centerline{\includegraphics[scale=0.5]{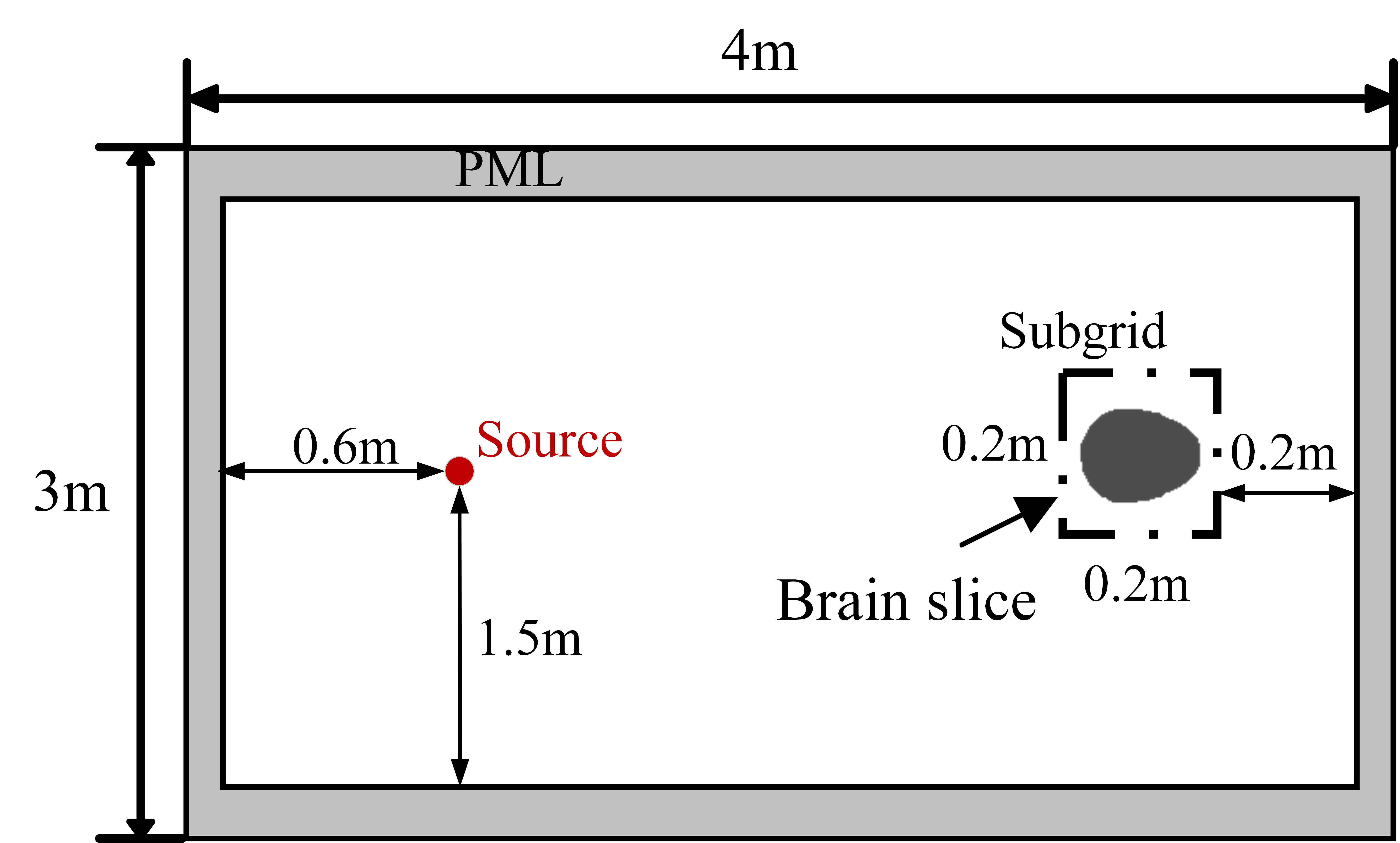}}  	
  	\caption{Structure of domain with human head cross-section.}
  	\label{CONFIGHUM}	
  \end{figure}

The last numerical example is the specific absorption
rate (SAR) calculation from a human head illuminated by an electromagnetic wave as a practical application to demonstrate the feasibility and the accuracy of the proposed FDTD subgridding method. The model used in the simulation is a cross-section of a human head from the computed tomography (CT) image, as shown in Fig. \ref{HUMANCT}. The gray values denote different materials. In our simulations, the detailed parameters of human tissues in Table I are obtained from \cite{ODOTP}. The computational domain is terminated with $10$-layered PML in Fig. \ref{CONFIGHUM}. The computational domain is $4$ m in length and $3$ m in width, and the human head is placed at $(3.6,1.5)$[m], as shown in Fig. \ref{CONFIGHUM}. The cell sizes of the coarse mesh are $4\times 10^{-3}$ m, and $2\times 10^{-3}$ m is used for the subgridding mesh region. The simulation time is $1\times 10^{-7}$ s.A $900$ MHz Gaussian pulse source is placed at (0.6, 1.5)[m] in the coarse gird.A time step of 4.6701 ps was chosen for the simulation in all the methods. The SAR is given by
 \begin{align}\label{CFL}   
 {\rm{SAR}} = \frac{{\sigma E_{{z_p}}^2}}{{2\rho }}, 
\end{align} 
where subscript $E_{{z_p}}$ denotes the peak absolute value of the electric field component. $\sigma $ and $\rho $ denote the specific conductance and density of the corresponding tissue respectively. According to \cite{ADSTF}, the surface integral is calculated as a measure of the accuracy 
\begin{align}
 {\left. {\mathop \sum \limits_i \mathop \sum \limits_j  {\rm{SAR}}} \right|_{i + \frac{1}{2},j + \frac{1}{2}}}{\rm{\Delta }}x{\rm{\Delta }}y. 
\end{align}
  
\begin{figure} 
	\begin{minipage}[h]{0.48\linewidth}\label{twlvea}
		\centerline{\includegraphics[scale=0.11]{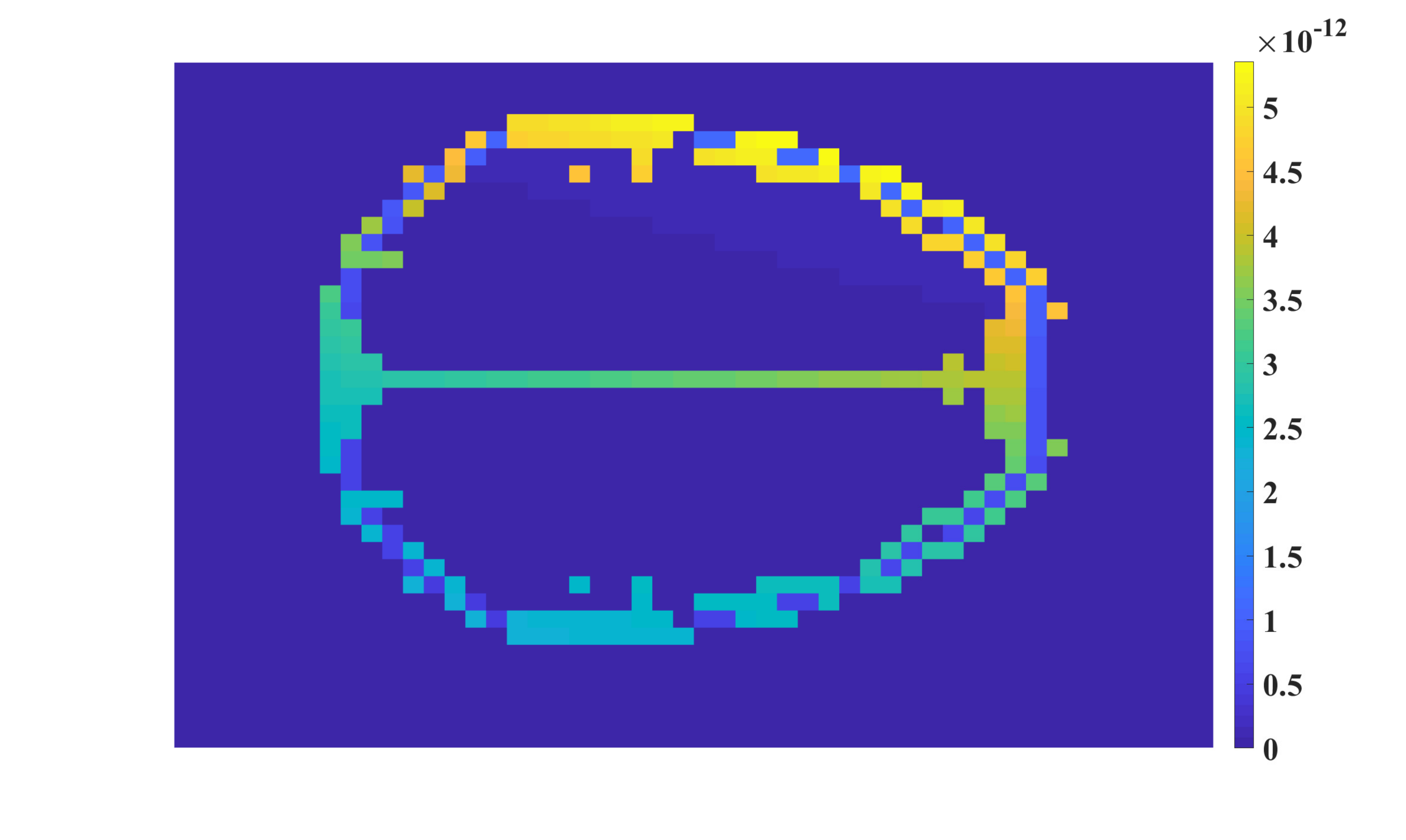}}
		\centerline{(a)}
	\end{minipage}
	\centering
	\begin{minipage}[h]{0.48\linewidth}\label{twlveb}
		\centerline{\includegraphics[scale=0.11]{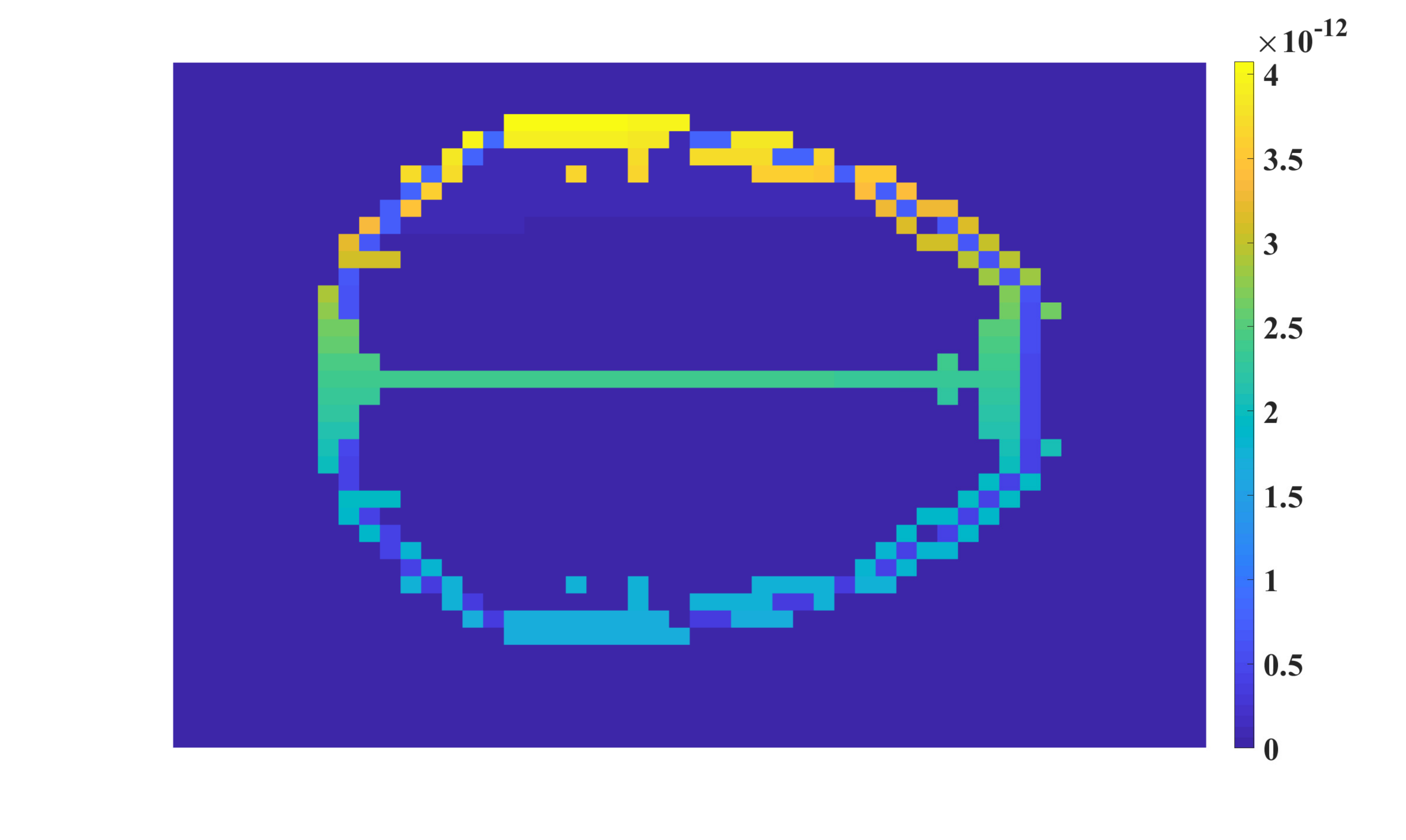}}
		\centerline{(b)}
	\end{minipage}
	\centering
		\begin{minipage}[h]{0.48\linewidth}\label{twlvec}
		\centerline{\includegraphics[scale=0.11]{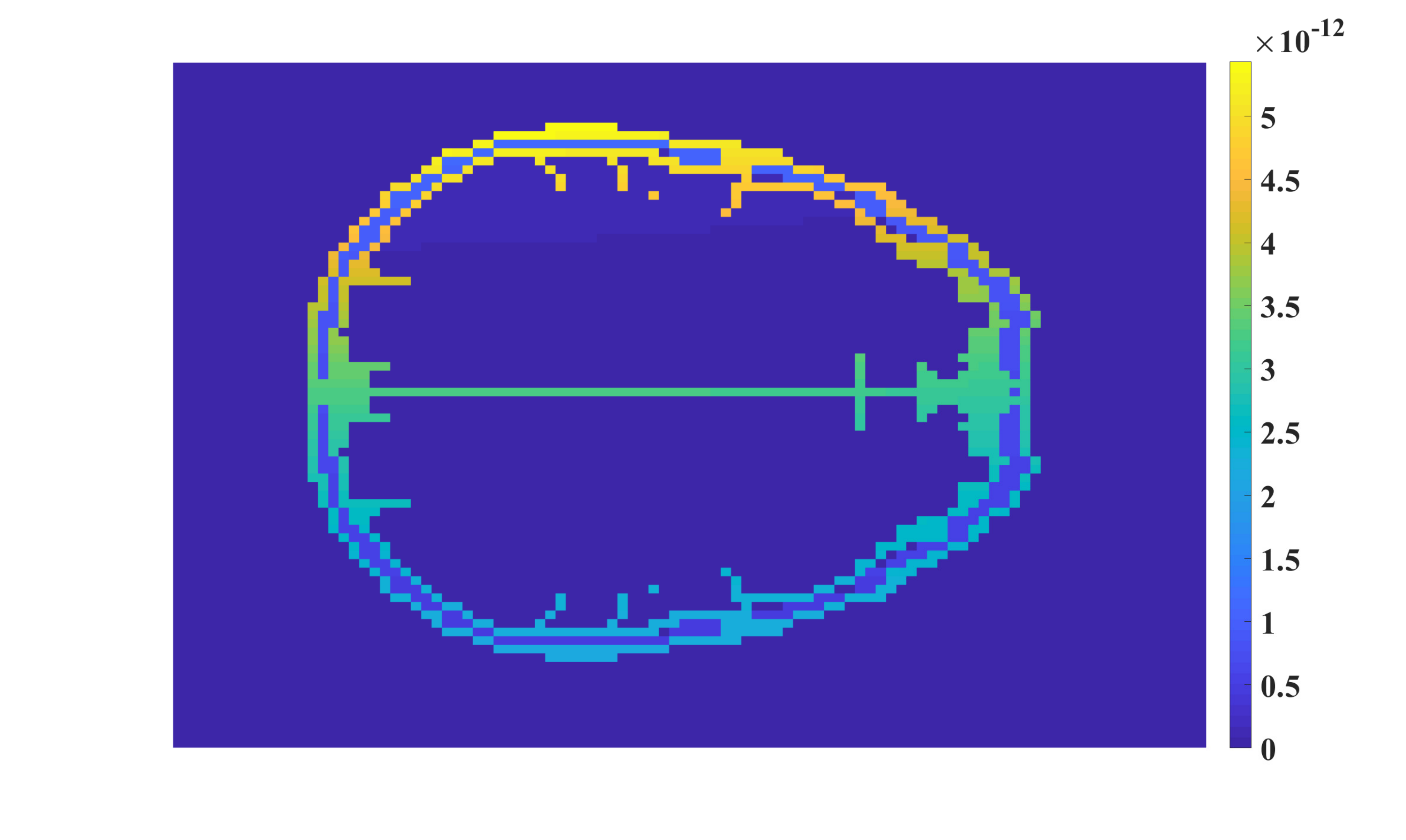}}
		\centerline{(c)}
	\end{minipage}
	\centering
	\begin{minipage}[h]{0.48\linewidth}\label{twlved}
		\centerline{\includegraphics[scale=0.11]{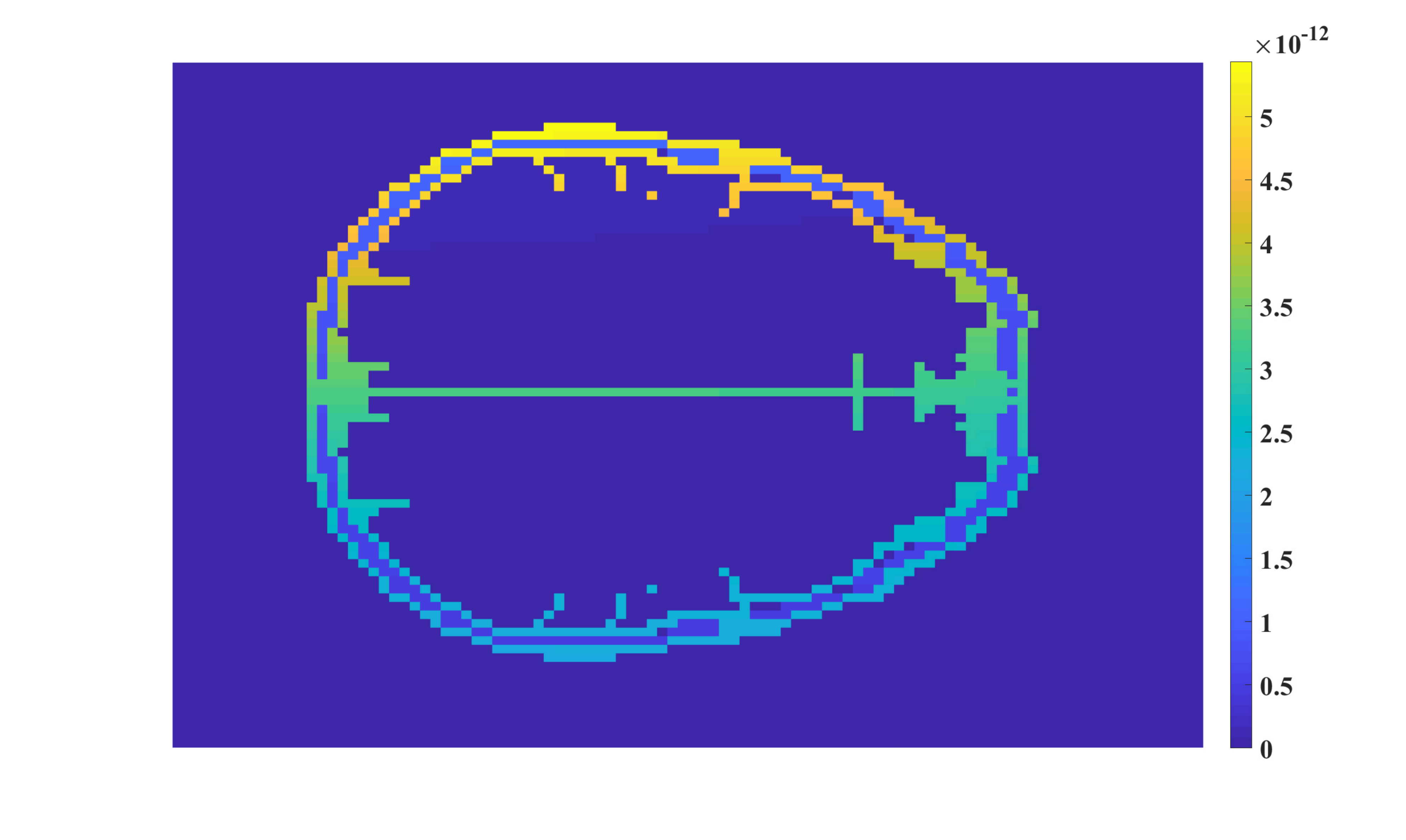}}
		\centerline{(d)}																																				
	\end{minipage}	
	\begin{minipage}[h]{0.48\linewidth}\label{twlvee}
	\centerline{\includegraphics[scale=0.08]{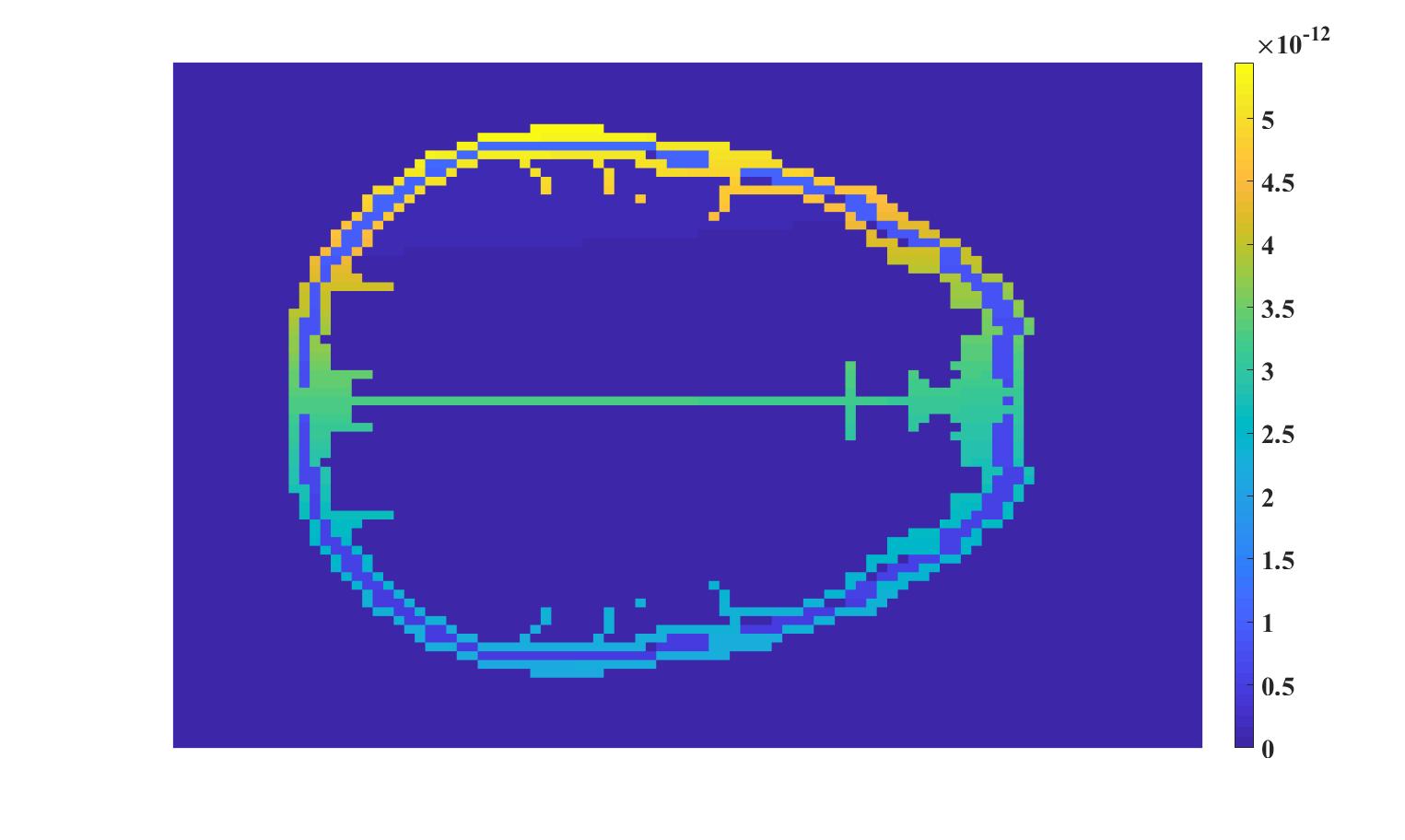}}
	\centerline{(e)}
\end{minipage} 
	\centering
	\caption{(a) The SAR obtained from the FDTD method with coarse meshes, (b) the SBP-SAT FDTD method with coarse meshes, (c) the FDTD method with fine meshes, (d) the SBP-SAT FDTD method with fine meshes, and (e) the proposed FDTD subgridding method.}
	\label{SARRES}
\end{figure}

Fig. \ref{SARRES} shows SAR obtained from the FDTD method, the SBP-SAT FDTD method and the proposed subgridding method. As shown in Fig. \ref{SARRES}(a) and (b), which show quite similar patterns,  many details of SAR are not shown since coarse meshes were used in the FDTD method and the SBP-SAT FDTD method. As fine meshes are used in the two methods, significantly more details are shown in Fig. \ref{SARRES}(c) and (d). Fig. \ref{SARRES}(e) shows the SAR calculated from the proposed SBP-SAT FDTD subgridding method. There are no visible differences between Fig. \ref{SARRES}(e) and Fig. \ref{SARRES}(c) and (d). Therefore, the proposed subgridding method can significantly improve the accuracy of the FDTD methods with coarse meshes.

Table II shows the computational consumption including relative errors of SAR obtained from different methods, the overall number of cells, and time costs. It can be found that there are 750,000 cells in coarse meshes, and 3,000,000 cells are used in the fine meshes. While there is only a slight increase in the number of cells (757,500) in the subgridding meshes. As shown in Table II, both the all-fine FDTD method and the SBP-SAT FDTD method with fine meshes can obtain accurate results, and the CPU time for the two methods are 5410.9s and 8,640.8s, respectively. Compared with those with coarse meshes, methods with fine meshes obviously use more CPU time. It should be noted that more CPU time is used in the SBP-SAT FDTD method compared with that of the FDTD method due to the overhead in the computation of the SAT terms. For the proposed subgridding method, however, since fine meshes are only used in a small region containing the human head model, the overall number of cells is 757,500, leading to only a 1.0\% increase of cells compared with coarse meshes. Meanwhile, only 0.42\% relative error exists in the proposed subgridding method and 2.4$\times$ speedup can be obtained, proving the local refinement’s availability.

\begin{table}
	\centering
	\caption{Computational consumption  of the FDTD method, the SBP-SAT FDTD method, the proposed SBP-SAT FDTD method }\label{table2}
	\resizebox{8.75cm}{!}{
	\begin{threeparttable}[b]
	\begin{tabular}{c|c|c|c|c}
		\hline
		\hline
		\textbf{Method} &\textbf{No. of Cells}   & \textbf{Relative Error} & {\textbf{Time Cost}} {[s]}  & {\textbf{Ratio}*}\cr 
		\hline
		\hline
		 FDTD fine meshes       &3,000,000 & -         &5,410.9  & -    \cr
		\hline
		 FDTD coarse meshes      &750,000   & 16.26$\%$ & 1,356.6 & 4.0   \\
		\hline
	    SBP-SAT FDTD fine meshes &3,000,000 &0.29$\%$   & 8,640.8 & 0.6 \\
		\hline
        SBP-SAT FDTD coarse meshes&750,000   &-16.72$\%$  & 2,123.9 &2.5   \\
	    \hline
	    Subgridding               &757,500    & 0.42$\%$   & 2,252.5 & 2.4  \\
		\hline 
		\hline 
	\end{tabular}
\begin{tablenotes}
	\footnotesize
	\item[*]Ratio is defined as the ratio of time cost in the FDTD method with fine meshes to that in the corresponding method.
\end{tablenotes}
\end{threeparttable}
}
\end{table}

\section{CONCLUSION}  
We proposed a stable FDTD subgridding method by combining the SBP operator and the SATs with the FDTD method to model local geometrically fine structures. By properly adding the electric field nodes in Yee's grid, the discrete operators can satisfy the SBP property. Therefore, the energy in the computational domain is fully determined by the fields on the boundaries. The boundary conditions for the computational domain and the interfaces between multiple mesh blocks are weakly enforced through the SATs. In addition, the interpolation matrices between different mesh blocks and parameters ensuring the stability are also derived. The staggered grids used in the SBP-SAT FDTD method are almost the same as those of the FDTD method. Only a few modifications are required when incorporating the proposed methods into the existing FDTD codes. 

Currently, an extension of the works in this paper into the three-dimensional space is in progress. We will report more results on this topic in the future.

\section*{APPENDIX}  
In this appendix, we listed all the detailed matrices used in the SBP-SAT FDTD method and the SBP-SAT FDTD subgridding method. The discrete partial differential operators are given by
\begin{align}
 &{\mathbb{D}}_{+ } = \left[ {\begin{array}{*{20}{r}}
 		{ - 1}&{\frac{1}{2}}&{\frac{1}{2}}&{}&{}&{}&{}\\
 		{ - \frac{1}{2}}&{ - \frac{1}{4}}&{\frac{3}{4}}&{}&{}&{}&{}\\
 		{}&{}&{ - 1}&1&{}&{}&{}\\
 		{}&{}&{}& \ddots & \ddots &{}&{}\\
 		{}&{}&{}&{}&{ - 1}&1&{}\\
 		{}&{}&{}&{}&{ - \frac{3}{4}}&{\frac{1}{4}}&{\frac{1}{2}}\\
 		{}&{}&{}&{}&{ - \frac{1}{2}}&{ - \frac{1}{2}}&1
 \end{array}} \right],  \\	 
 &{\mathbb{D}}_{- }= \left[ {\begin{array}{*{20}{r}}
 	{ - 1}&1&{}&{}&{}&{}&{}&{}\\
 	{ - 1}&1&{}&{}&{}&{}&{}&{}\\
 	{ - \frac{1}{5}}&{ - \frac{3}{5}}&{\frac{4}{5}}&{}&{}&{}&{}&{}\\
 	{}&{}&{ - 1}&1&{}&{}&{}&{}\\
 	{}&{}&{}& \ddots & \ddots &{}&{}&{}\\
 	{}&{}&{}&{}&{ - 1}&1&{}&{}\\
 	{}&{}&{}&{}&{}&{ - \frac{4}{5}}&{\frac{3}{5}}&{\frac{1}{5}}\\
 	{}&{}&{}&{}&{}&{}&{ - 1}&1\\
 	{}&{}&{}&{}&{}&{}&{ - 1}&1
 	\end{array}} \right]. 
 \end{align}
 
${\mathbb{P}}_{+ }$, ${\mathbb{P}}_{-}$, ${\mathbb{Q}}_{+ }$ and ${\mathbb{P}}_{-}$ are given by
 
\begin{align} 
&{\mathbb{P}}_{+ } = diag\left( {\left[ {\frac{1}{2},1,1,...,1,\frac{1}{2}} \right]} \right)h, \\
&{\mathbb{P}}_{- } = diag\left( {\left[ {\frac{1}{2},\frac{1}{4},\frac{5}{4},1,...,1,\frac{5}{4},\frac{1}{4},\frac{1}{2}} \right]} \right)h, 
\end{align}

\begin{align}
&{\mathbb{Q}}_{+ } = \left[ {\begin{array}{*{20}{r}}
		{ - \frac{1}{2}}&{\frac{1}{4}}&{\frac{1}{4}}&{}&{}&{}&{}\\
		{ - \frac{1}{2}}&{ - \frac{1}{4}}&{\frac{3}{4}}&{}&{}&{}&{}\\
		{}&{}&{ - 1}&1&{}&{}&{}\\
		{}&{}&{}& \ddots & \ddots &{}&{}\\
		{}&{}&{}&{}&{ - 1}&1&{}\\
		{}&{}&{}&{}&{ - \frac{3}{4}}&{\frac{1}{4}}&{\frac{1}{2}}\\
		{}&{}&{}&{}&{ - \frac{1}{4}}&{ - \frac{1}{4}}&{\frac{1}{2}}
\end{array}} \right],  \\
&{\mathbb{Q}}_{- } = \left[ {\begin{array}{*{20}{r}}
	{ - \frac{1}{2}}&{\frac{1}{2}}&{}&{}&{}&{}&{}&{}\\
	{ - \frac{1}{4}}&{\frac{1}{4}}&{}&{}&{}&{}&{}&{}\\
	{ - \frac{1}{4}}&{ - \frac{3}{4}}&1&{}&{}&{}&{}&{}\\
	{}&{}&{ - 1}&1&{}&{}&{}&{}\\
	{}&{}&{}& \ddots & \ddots &{}&{}&{}\\
	{}&{}&{}&{}&{ - 1}&1&{}&{}\\
	{}&{}&{}&{}&{}&{ - 1}&{\frac{3}{4}}&{\frac{1}{4}}\\
	{}&{}&{}&{}&{}&{}&{ - \frac{1}{4}}&{\frac{1}{4}}\\
	{}&{}&{}&{}&{}&{}&{ - \frac{1}{2}}&{\frac{1}{2}}
	\end{array}} \right]. 
\end{align}
In addition, the upper left corner of ${{\mathbb{T}}_1}$ is given by 
\begin{align}
\begin{array}{l}
{a_{1,1}} = 0.3405,{a_{1,2}} = 0.1703,{a_{1,3}} = 0.5487,\\
{a_{1,4}} = 0.1126,{a_{1,5}} =  - 0.0541,{a_{1,6}} =  - 0.1180,\\
{a_{2,1}} = 0.3405,{a_{2,2}} = 0.1703,{a_{2,3}} = 0.5487,\\
{a_{2,4}} = 0.1126,{a_{2,5}} =  - 0.0541,{a_{2,6}} =  - 0.1180,\\
{a_{3,1}} =  - 0.0043,{a_{3,2}} =  - 0.0022,{a_{3,3}} = 0.1708,\\
{a_{3,4}} = 0.3324,{a_{3,5}} =  - 0.3324,{a_{3,6}} = 0.1708,\\
{a_1} = 0.375,{a_2} = 0.125. 
\end{array} 
\end{align}

\end{document}